\documentclass[sigconf]{acmart}

\AtBeginDocument{%
  \providecommand\BibTeX{{%
    \normalfont B\kern-0.5em{\scshape i\kern-0.25em b}\kern-0.8em\TeX}}}

\copyrightyear{2024}
\acmYear{2024}
\setcopyright{acmlicensed}\acmConference[CHI '24]{Proceedings of the CHI Conference on Human Factors in Computing Systems}{May 11--16, 2024}{Honolulu, HI, USA}
\acmBooktitle{Proceedings of the CHI Conference on Human Factors in Computing Systems (CHI '24), May 11--16, 2024, Honolulu, HI, USA}
\acmDOI{10.1145/3613904.3642803}
\acmISBN{979-8-4007-0330-0/24/05}
\newcommand{\tool}{\textsc{PromptCharm}}

\newcommand{\circled}[1]{{\large \textcircled{\normalsize #1}}}

\newcommand{\rcolorbox}[2]{{#2}
}

\usepackage{xcolor}

\definecolor{primary}{HTML}{3f51b5}
\definecolor{secondary}{HTML}{f44336}

\definecolor{lightgray}{HTML}{eeeeee}
\definecolor{tab_red}{rgb}{1,0.76,0.71}
\definecolor{tab_purple}{HTML}{FDD0D0}
\definecolor{t_test_highlight}{HTML}{FFEBEE}
\definecolor{icongray}{HTML}{767576}

\usepackage{multirow}
\newcommand\promptsearchicon{\includegraphics[scale=0.15]{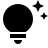}}

\newcommand\promptaddicon{\includegraphics[scale=0.15]{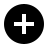}}

\usepackage{pifont} 

\usepackage{xspace}
\usepackage[ruled,linesnumbered]{algorithm2e}    % for algorithm block
\usepackage{colortbl}
\usepackage{subcaption}
\usepackage{enumitem}

\makeatletter
\DeclareRobustCommand\onedot{\futurelet\@let@token\@onedot}
\def\@onedot{\ifx\@let@token.\else.\null\fi\xspace}

\makeatother

\newcommand{\responseref}[0]{\color{black}}
\newcommand{\responseline}[1]{\textcolor{black}{#1}}
\newcommand{\camerareadyrevision}[1]{\textcolor{black}{#1}}
\newcommand{\smalltitle}[1]{\vspace{1mm}{\noindent\bf \textit{#1}}}

\begin{document}

%\title{\tool : Visualizing, Understanding, and Debugging Recurrent Neural Networks}
\title{{\tool}: Text-to-Image Generation through Multi-modal Prompting and Refinement}
% \title{{\tool}: Text-to-Image Creation through Interactive and Multi-Perspective Prompt Engineering\todo{paper title can be more exciting. "text-to-image creation" sounds a bit boring. "multi-perspective" sounds like a new word and is a bit vague here. "multi-modal prompting"? I hope you can also highlight "refinement" or "enriching feedback loop to Diffusion models" in the title.}}

\author{Zhijie Wang}
\affiliation{%
  \institution{University of Alberta}
  \city{Edmonton}
  \state{AB}
  \country{Canada}
}
\email{zhijie.wang@ualberta.ca}

\author{Yuheng Huang}
\affiliation{%
  \institution{University of Alberta}
  \city{Edmonton}
  \state{AB}
  \country{Canada}}
\email{yuheng18@ualberta.ca}

\author{Da Song}
\affiliation{%
  \institution{University of Alberta}
  \city{Edmonton}
  \state{AB}
  \country{Canada}}
\email{dsong4@ualberta.ca}

\author{Lei Ma}
% \authornote{Lei Ma is also affiliated with Alberta Machine Intelligence Institute (Amii), Canada.}
\affiliation{%
  \institution{The University of Tokyo, Japan}
  \institution{University of Alberta, Canada}
  \city{}
  \state{}
  \country{}
  }
\email{ma.lei@acm.org}

\author{Tianyi Zhang}
\affiliation{%
  \institution{Purdue University}
  \city{West Lafayette}
  \state{IN}
  \country{USA}}
\email{tianyi@purdue.edu}

% \thanks{Zhijie Wang, Yuheng Huang, Da Song, and Lei Ma are also affiliated with the Alberta Machine Intelligence Institute (Amii), Canada.}
\begin{abstract}
The recent advancements in Generative AI have significantly advanced the field of text-to-image generation. The state-of-the-art text-to-image model, \textit{Stable Diffusion}, is now capable of synthesizing high-quality images with a strong sense of aesthetics. Crafting text prompts that align with the model's interpretation and the user's intent thus becomes crucial. However, prompting remains challenging for novice users due to the complexity of the stable diffusion model and the non-trivial efforts required for iteratively editing and refining the text prompts. To address these challenges, we propose {\tool}, a mixed-initiative system that facilitates text-to-image creation through multi-modal prompt engineering and refinement. 
% \responseline{
To assist novice users in prompting, {\tool} first automatically refines and optimizes the user's initial prompt. Furthermore, {\tool} supports the user in exploring and selecting different image styles within a large database. To assist users in effectively refining their prompts and images, {\tool} renders model explanations by visualizing the model's attention values. If the user notices any unsatisfactory areas in the generated images, they can further refine the images through model attention adjustment or image inpainting within the rich feedback loop of {\tool}. To evaluate the effectiveness and usability of {\tool}, we conducted a controlled user study with 12 participants and an exploratory user study with another 12 participants.
These two studies show that participants using {\tool} were able to create images with higher quality and better aligned with the user's expectations compared with using two variants of {\tool} that lacked interaction or visualization support.
% }
% \todo{rewrite the following sentences to make it more obvious what are the multiple modalities.}To provide novice users \todo{end-user or novice user? In HCI, these two terms have different meanings. Keep it consistent.} a starting point, {\tool} automatically refine and optimize the user's initial text prompt. To enrich the feedback loop in prompt engineering, {\tool} supports \todo{your HCI reviewers may not get what is a modifier in this context.}modifier exploration, model attention adjustment, and \todo{simply call it image impainting without any context may leave an impression that this is a simple photoshop technique.}image inpainting. Lastly, {\tool} provides \todo{try not to call this "rich XAI feedback" since it is only one kind of XAI feedback---model attention.}rich explainable AI (XAI) feedback through model attention visualization to help users iteratively refine their text prompts and improve their creations. To evaluate the effectiveness and usability of {\tool}, we conducted a controlled user study with \todo{X} participants and an exploratory user study with another \todo{X} participants. 
% % \todo{you need to be a bit specific here about the nature of these two studies. You can say "we conducted a controlled study and also an exploratory study..."}
% These two studies show that participants using {\tool} were able to create images with higher quality and better visual effects while experiencing less mental demand compared with using the variants of {\tool} that lacked interaction or visualization support.
\end{abstract}

\begin{CCSXML}
<ccs2012>
   <concept>
       <concept_id>10003120.10003121.10003129</concept_id>
       <concept_desc>Human-centered computing~Interactive systems and tools</concept_desc>
       <concept_significance>500</concept_significance>
       </concept>
   <concept>
       <concept_id>10010147.10010257</concept_id>
       <concept_desc>Computing methodologies~Machine learning</concept_desc>
       <concept_significance>500</concept_significance>
       </concept>
 </ccs2012>
\end{CCSXML}

\ccsdesc[500]{Human-centered computing~Interactive systems and tools}
\ccsdesc[500]{Computing methodologies~Machine learning}

\keywords{Generative AI, Prompt Engineering, Large Language Models}

\begin{teaserfigure}
  \centering
  \includegraphics[width=\linewidth]{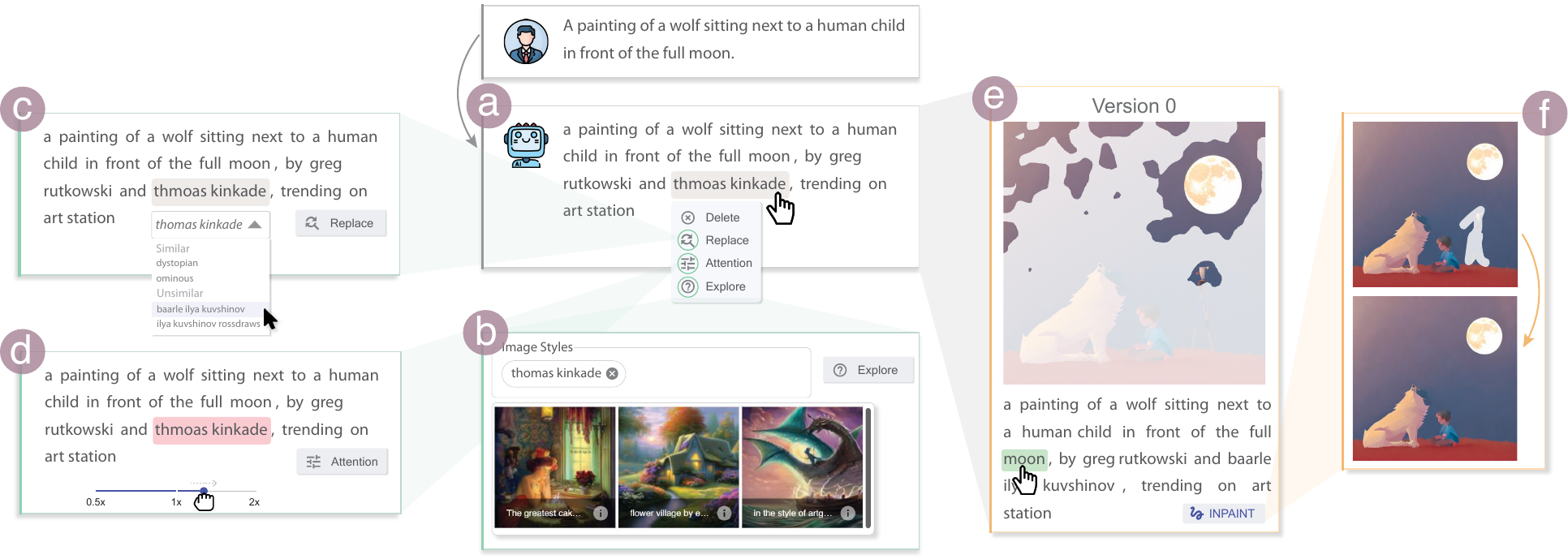}
  \caption{\textbf{{\tool} facilitates prompt engineering in text-to-image generation with an enriched, multi-modal feedback loop.}
  %multi-granularity \todo{"multi-granularity" sounds strange here. Maybe "multi-modal"?}prompt engineering and image editing.} 
  (a) Given an initial prompt from a user, {\tool} first suggests an initial refinement based on a prompt optimization model. 
  % To enrich the feedback loop to a stable diffusion model. 
  %To help users \todo{I think it's less about gaining control. It's more about enriching the feedback loop to a diffusion model.}gain more control over the generation, 
  (b) The user can explore different styles by searching them in a large database. (c) The user can also explore similar and dissimilar image styles in a drop-down list. (d) Furthermore, the user can adjust the attention of the diffusion model to different keywords in the prompt. (e) Given an initial image generated by the model, the user can examine which part of the image corresponds to which part of the text prompt. (f) The user can further mark undesired parts of an image to remove or regenerate them through an image inpainting model.
  %{\tool} further supports (b) modifier replacement, (c) attention adjustment, (d) modifier exploration, and (f) image inpainting. Moreover, to assist users iteratively improve their creation, {\tool} provides rich XAI feedback via model attention visualization for the generated image (e). 
  %Finally, {\tool} also supports direct image editing through model-driven inpainting (f).\todo{inpainting sounds a bit isolated from other features. Why don't categorize it also into a feedback modality from users to diffusion models?} 
  }
  \Description{This figure shows the seven core features of PromptCharm. (a) The user can first type their initial input prompt in a text box. PromptCharm will then automatically refine the user's prompt. (b) The user can explore different styles by typing in a search bar and searching them in a large database. (c) The user can also explore similar and dissimilar image styles in a drop-down list. (d) Furthermore, the user can adjust the attention of the diffusion model to different keywords in the prompt by controlling a slider. (e) Given an initial image generated by the model, the user can examine which part of the image corresponds to which part of the text prompt. (f) The user can further mark undesired parts of an image with a brush to remove or regenerate them through an image inpainting model.}
  \label{fig:teaser_ui}
\end{teaserfigure}

\sloppy 
\maketitle

\section{Introduction}
\label{sec:intro}

%% Challenges
% 1. writing prompts is difficult for novice users.
% 2. it is hard for end-users to interpret model's generation results
% 3. users lack of control over the generation process

% \todo{merge the first two sentences to a short sentence. they read a bit verbose}
The recent advancements in Generative AI have brought significant progress to text-to-image generation---an intersection field of computer vision (CV) and natural language processing (NLP).
% Generating high-quality images based on text input has posed a longstanding challenge at the intersection field of computer vision (CV) and natural language processing (NLP). The recent advancements in Generative AI have brought significant progress to this field. 
State-of-the-art (SOTA) text-to-image models such as \textit{Stable Diffusion}~\cite{rombach2022high} and \textit{DALL-E}~\cite{openai2021dalle} have showcased impressive capabilities in producing images with exceptional quality and fidelity. As a result, these text-to-image models find utility across diverse domains, including visual art creation~\cite{ko2023large}, news illustration~\cite{liu2022opal}, and industrial design~\cite{liu20233dall}. According to the recent studies~\cite{liu2022design,oppenlaender2022prompt}, the quality of AI-generated images is highly sensitive to the text prompts. 
%\todo{here misses a sentence to explain that the image quality is highly sensitive to text prompts.} 
Thus, crafting text prompts (also known as {\em prompting} or {\em prompt engineering}) has emerged as a crucial step in text-to-image generation. 

Previous studies have highlighted that novice users often struggle with writing prompts~\cite{zamfirescu2023johnny, ko2023large}. Specifically, novice users
%\todo{end-user or novice user? In HCI, these two terms have different meanings. Keep it consistent.} 
often experience a steep learning curve when attempting to write text prompts that the model can effectively interpret while preserving their creative intentions. Moreover, generating images with a profound sense of aesthetics requires domain knowledge in creative design, particularly in employing specific modifiers (i.e., \textit{magic words} about image styles)~\cite{weisz2023toward, liu2022design}. Unfortunately, novice users often do not have such expertise. 

Recently, several interactive approaches have been proposed to support prompt engineering for natural language processing~\cite{strobelt2022interactive} or computer vision tasks~\cite{wang2023reprompt, brade2023promptify, feng2023promptmagician}.
% \todo{interactive? if so, say "interactive techniques". This will make your work sound more related to HCI.} techniques for prompt engineering~\cite{weisz2023toward, liu2022design, wang2023reprompt, zamfirescu2023johnny, wang2022diffusiondb}. These investigations have yielded valuable insights into, e.g., how to formulate a prompt, the importance of prompt modifiers, and the taxonomy of prompt modifiers. However, \todo{I don't quite get this sentence. Can you write it in a more straightforward way?}these high-level recommendations struggle to seamlessly align with the unique creative intentions of individual users. \todo{Oh I see. You start talking about interactive prompting techniques from here. Then why bother talking about those needfinding studies or empirical studies? You should go straight to interactive approaches in this paragraph and save those studies for your needfinding section. No need to expose them this early.}
% Specifically, a few recent techniques have been focusing on interactive support for prompt engineering~\cite{strobelt2022interactive, brade2023promptify}. 
These approaches aim to guide the iterative prompt refinement process by rendering a set of alternative prompts~\cite{strobelt2022interactive} or suggesting a few new keywords for users to choose from
% \todo{"automatically refine"? Sounds contradicting to the "interactive approaches" in the previous sentence.}automatically refining prompts
~\cite{brade2023promptify, feng2023promptmagician}. 
% \responseline{
Nevertheless, existing approaches usually lack rich feedback during the user's creation process. For instance, users may wonder to what extent the model has incorporated their text prompts during the generation. However, without appropriate support to explain the model's generation, the user may find it difficult to interpret a generated image. As a result, they do not know which part of their prompts has worked and which has not. Thus, they may feel clueless when refining their text prompts for the new iteration. %A trial-and-error process of prompting is inevitable before the user feels satisfied with their creations.
% }
% Nevertheless, existing techniques usually 
% %\todo{I like this phrase "lack rich feedback". I want you to highlight it more throughout the paper.}
% lack rich feedback on \todo{"each trail of the user's creation process" sounds strange and also verbose. I haven't seen such a narrative in any HCI paper.}each trail of the user's creation process. For instance, when a new modifier\todo{unclear what modifiers mean in this context. need to use a different word or give some explanation first.} is suggested, users may wonder how this modifier would possibly affect the generated image's visual effects. \todo{this example is not very impressive. A HCI reviewer may say that you can just add a simple preview. Replace it with a more impressive example. For example, I think the model attention visualization is more impressive than the current example.}Without appropriate support to help users explore these modifiers ahead of the generation, a trail-and-error process of prompting is inevitable before users feel satisfied with their creations.

% \todo{I like this paragragraph.}
The deficiency of lacking proper feedback becomes especially pronounced when confronted with the situation that the model's output does not align with a user's intention. For instance, a subject mentioned in the text prompt is missing in the generated image. As a result, it becomes arduous for the user to revise their prompt effectively. This is particularly due to the complexity of the stable diffusion model and the absence of suitable model explanations. Moreover, recent studies from the ML community have shown that SOTA generative AI and large language models could actually misinterpret the user's intention in the input text prompt~\cite{kou2023model, garcia2023uncurated}. A more tangible way to align the user's creative intentions with the model's generation becomes another urgent need in facilitating text-to-image creation. 

In this paper, we present 
%\todo{no need to put the name in bold.}
{\tool}, a mixed-initiative system that supports the iterative refinement of AI-generated images
%\todo{"user-generated content"? Shouldn't it be "AI-generated images". Also, since this paper is about image generation, better to use "image" instead of "content"} 
by enabling multi-modal prompt engineering within a rich feedback loop.
% \responseline{
Since novice users might have little experience in prompting, {\tool} leverages a SOTA prompt optimization model, \textit{Promptist}~\cite{hao2022optimizing}, to automatically revise and improve their initial input prompts. The user can then efficiently explore different image styles and pick modifiers they are interested in through {\tool}. The user can further examine which part of the generated image corresponds to which part of the text prompt by observing the model attention visualization in {\tool}. As the user notices a misalignment between the generated image and their input prompts, they can refine the generated image in {\tool} by adjusting the attention of the model to specific keywords in the given prompts. They can also mark undesired parts of an image to remove or regenerate them through an image inpainting model. With the help of {\tool}, the user can avoid re-writing their prompts to match the model's interpretation with their creative intent, as such revision may lead to a tedious process of \textit{trail-and-error}. Finally, {\tool} provides version control to help users easily track their image creations within an iterative process of prompting and refinement.
To evaluate the effectiveness and usability of {\tool}, we conducted two within-subjects user studies with a total of 24 participants who had no more than one year of experience in using text-to-image generative models. We created two variants of {\tool} as comparison baselines (denoted as \textit{Baseline} and \textit{Promptist}) by disabling novel interactions and features in {\tool}. In the first study with 12 participants featuring close-ended tasks, participants using {\tool} were able to create images with the highest similarity to the target images across all three tasks (average SSIM: $0.648\pm0.100$) compared with the participants using Baseline (average SSIM: $0.479\pm0.115$) or Promptist (average SSIM: $0.574\pm0.111$). In the second study with another 12 participants featuring open-ended tasks, participants self-reported more satisfied with their images when using {\tool} in terms of aesthetically pleasing compared with using Baseline ($5.8$ vs. $4.9$, \responseline{Wilcoxon signed-rank test: $p=0.02$}) and Promptist ($5.8$ vs. $4.9$, \responseline{Wilcoxon signed-rank test: $p=0.01$}) on a 7-point Likert scale.
Participants also felt their images matched their expectations better compared with using either Baseline ($5.9$ vs. $4.4$, \responseline{Wilcoxon signed-rank test: $p=0.02$}) or Promptist ($5.9$ vs. $4.8$, \responseline{Wilcoxon signed-rank test: $p=0.04$}).
% with a total of 24 participants\todo{"a total of 24 participants" may lead to a question about whether the same participant has participated into both studies. better describe them separately --- a study with X participants and another study with another set of Y participants. } with various levels of expertise in Generative AI. We created two variants of {\tool} as comparison baselines by disabling the novel interactions and features in {\tool}. In the first study featuring close-ended tasks, participants using \todo{the writing from here to the end of the paragraph reads a bit sloppy. Should rewrite with some placeholders for objective measures to make it read more scientific and rigorous}{\tool} were able to create images that look more similar to the given target images compared with the participants using two baselines. In the second study involving open-ended tasks, we \todo{to finish after the user study.}. 
These results demonstrate that {\tool} can assist users in effectively creating images with higher quality and senses of aesthetics while not requiring rich relevant experience. 
% }

In summary, this paper makes the following contributions:

\begin{itemize}
    \item {\tool}, a mixed-initiative interaction system that supports text-to-image creation through multi-modal prompting and image refinement for novice users. We have open-sourced our system on Github~\footnote{https://github.com/ma-labo/PromptCharm}.
    
    \item A set of visualizations, interaction designs, and implementations for interactive prompt engineering.
    
    \item Two within-subjects user studies to demonstrate that {\tool} facilitates users to create better and more aesthetically pleasing images compared with two baseline tools.
\end{itemize}

\section{Related Work}
\label{sec:related_work}

\subsection{Text-to-Image Generation}
\label{subsec:text_to_image}
% \todo{This subsection should be 2.1.}
Text-to-image generation stands as a pivotal capability within the realm of generative AI. Given a text input, an AI model aims to generate an image whose content is aligned with the text description. One of the pioneering attempts in the field of text-to-image generation is AlignDraw~\cite{mansimov2015generating}. AlignDraw is extended from Draw~\cite{gregor2015draw}, an RNN-based image generation model, by leveraging a bidirectional attention RNN language model to guide the image generation process. With the advancements in generative adversarial networks (GANs)~\cite{goodfellow2014generative}, a large body of research on text-to-image generation has been focusing on GAN-based approaches~\cite{reed2016generative, zhang2017stackgan, li2019controllable, esser2021taming}. Reed et al. proposed one of the earliest GAN-based text-to-image generation models by combining GAN with a convolutional-recurrent text encoder~\cite{reed2016generative}. To improve the resolution and quality of the generated images, Zhang et al. then utilized a two-stage model architecture, StackGAN, to gradually synthesize and refine an image~\cite{zhang2017stackgan}. The emergence of the transformer~\cite{vaswani2017attention} models further enhanced the GAN model's capability of generating high-quality images~\cite{esser2021taming}.

Notably, the success of the transformer models in a wide range of NLP tasks has also inspired computer vision researchers. The transformer-based architecture later showed promising performance on text-to-image generation~\cite{openai2021dalle, ding2021cogview, wu2022nuwa}. In 2021, Open AI released DALL-E, a GPT-3-based model for text-to-image generation~\cite{openai2021dalle}. Microsoft proposed NUWA, a pre-trained 3D encoder-decoder transformer for various visual synthesis tasks, including text-to-image generation~\cite{wu2022nuwa}. Recently, denoising diffusion probabilistic models (DDPM, usually also referred as diffusion models)  have surged as the dominant approach in text-to-image generation research~\cite{gu2022vector, openai2022dalle2, rombach2022high}. Compared with GAN-based and transformer-based methods, diffusion models are capable of generating images with much higher resolutions. CLIP (contrast language-image pre-training) further enhances the diffusion model's ability to understand both linguistic and visual concepts~\cite{radford2021learning}. As a result, SOTA text-to-image generation models (e.g., Stable Diffusion~\cite{rombach2022high} and DALL-E2\cite{openai2022dalle2}) are mostly based on a pipeline of combining diffusion models with CLIP. 
%\todo{You should end with Stable Diffusion instead of CLIP. Otherwise it feels CLIP is the state of the art and your reviewers will ask you why not use CLIP.} 
The Stable Diffusion model~\cite{rombach2022high} was trained on the LAION dataset~\cite{schuhmann2022laion} with latent diffusion models and cross-attention layers. {\tool} leverages the Stable Diffusion model~\cite{rombach2022high} as the text-to-image generation pipeline since it is open-sourced and has SOTA performance on public benchmarks.

\subsection{Prompt Engineering}
\label{subsec:prompt_engineering}

% 1. general prompt engineering
% prompting tricks: few-shot prompting
% brown2020language gao2021making zhao2021calibrate
%
% prompting tricks: chain-of-thoughts
% wei2022chain yao2023 wu2022ai
%
% prompting guidelines
% liu2022makes zamfirescu2023johnny liu2022design oppenlaender2022prompt
%
% automated prompting 
% shin2020autoprompt wen2023hard pavlichenko2023best reprompt 
%
% 2. interactive prompt engineering
% brade2023promptify strobelt2022interactive feng2023promptmagician
% \todo{merge the first two sentences. Reads verbose.}
% \responseline{
The widespread use of generative models has raised the significance of prompting. As a result, an increasing number of techniques have been proposed for prompt engineering. For instance, few-shot prompting~\cite{brown2020language, gao2021making, zhao2021calibrate} and chain-of-thought prompting~\cite{wei2022chain, yao2023tree, wu2022ai} are both representative prompting techniques for generative language models.
% One of the representative prompting techniques---few-shot prompting---requires users to add instructions with several examples in the prompt~\cite{brown2020language, gao2021making, zhao2021calibrate}. Therefore, a pre-trained generative model can understand unseen or personalized user tasks through in-context learning. Another representative prompting technique is the chain-of-thought prompting~\cite{wei2022chain, yao2023tree, wu2022ai}, which instructs a large language model to solve a problem \textit{step-by-step} by adding specific instructions, e.g., ``\textit{let's think step-by-step}'' in the prompt. 
In addition to prompting techniques, there has also been a few studies about prompting guidelines~\cite{liu2022design, zamfirescu2023johnny, liu2022makes, oppenlaender2022prompt}. To identify the prompts that can effectively help text-to-image models generate coherent outputs, Liu et al.~conducted user studies with practitioners to derive a set of prompt design guidelines~\cite{liu2022design}. They specifically found that style keywords (modifiers) play a vital role in affecting the generated image's quality. This is also confirmed by a recent study on the taxonomy of prompt modifiers for text-to-image generation~\cite{oppenlaender2022prompt}. 

Since hand-crafting prompts require significant manual efforts, another line of research has focused on automated prompt generation~\cite{shin2020autoprompt, wen2023hard, pavlichenko2023best, wang2023reprompt, hao2022optimizing}. %Shin et al.~proposed AutoPrompt, an automated prompting method for large language models based on a gradient-guided search strategy~\cite{shin2020autoprompt}. Wang et al.~developed RePrompt, a prompt refining pipeline that helps improve the emotional expressions in a prompt~\cite{wang2023reprompt}. 
Specifically, for text-to-image generation, recent research efforts have been focusing on automatically optimizing user input prompts and extending them with effective image style keywords (modifiers)~\cite{hao2022optimizing, wen2023hard, pavlichenko2023best}. For instance, Wen et al.~proposed a method to learn prompts that can be re-used across different image generation models through gradient-based discrete optimization. Notably, our work complements this line of research, since {\tool} can be combined with any automated prompting methods for text-to-image generative models. We specifically select Promptist~\cite{hao2022optimizing}, a reinforcement learning-based method, as our prompt refinement model given its superior performance.

Our work is most related to the interactive prompt engineering~\cite{brade2023promptify, strobelt2022interactive, feng2023promptmagician}. 
% \todo{try to condense the following two sentences into one sentence}
Strobelt et al.~proposed PromptIDE, an interactive user interface to help users explore different prompting options for NLP tasks~\cite{strobelt2022interactive}. To assist novice users in prompting for text-to-image generation, Promptify provides an interactive user interface that allows users to explore different generated images and iteratively refine their prompts based on the suggestions from a GPT3 model~\cite{brade2023promptify}. Another recent work, PromptMagician, utilizes an image browser to support users to efficiently explore and compare the generated images with images retrieved from a database~\cite{feng2023promptmagician}. \responseline{Both Promptify and PromptMagician aim at assisting users in exploring a large set of different generated images when revising text prompts. Different from them, {\tool} focuses on helping users iteratively improve one generated image through multi-modal prompting by adjusting the model's attention to keywords in the prompt. By adjusting the model's attention, the user does not need to rewrite their prompts to align the model's interpretation with their creative intention. Therefore, they can avoid risking completely changing the image content when revising the prompt. Another recent work, PromptPaint, also supports prompting beyond text by providing flexible steering through paint medium-like interactions~\cite{chung2023promptpaint}. \camerareadyrevision{By masking undesired areas and directly inpaint them, the user could efficiently remove or regenerate some areas of an image while preserving the other areas. Different from PromptPaint, {\tool} further provides model explanations to help users interpret the model's generation.} \camerareadyrevision{Subsequently, users can improve the generated images by enhancing specific parts of images or prompts identified by observing model explanations.}}
% \responseline{Different from the aforementioned literature, PromptPaint not only supports users in discovering prompt options, but also provides flexible steering through paint medium-like interactions~\cite{chung2023promptpaint}}
%Therefore, users can gain insights from exploration and gradually improve their prompts. 
% \responseline{
% Our work differs from these systems in two ways. First, {\tool}  supports multi-modal prompting by adjusting the model's attention to keywords in the prompt and image inpainting. By adjusting the model's attention, the user does not need to re-write their prompts to align the model's interpretation with their creative intention. Therefore, they can avoid risking completely changing the image content when revising the prompt. By masking undesired areas and directly inpaint them, the user could efficiently remove or regenerate some areas of an image while preserving the other areas.
% \todo{need to add more elaborations on why adjusting model attention and image inpainting is better than other methods} \todo{the following sentence reads a bit short and shallow.}Thus, when the user's intention is not aligned with the generative model's interpretation, {\tool} can efficiently address this. 

\responseline{Note that open-sourced tools outside of academic research such as \textit{Stable Diffusion Web UI} (SD Web UI)~\cite{sdwebui} also provides similar features such as attention adjustment and visualization. However, {\tool} is different from SD Web UI in the following ways. 
First, SD Web UI only provides attention visualization over the image (such as a heatmap). By contrast, {\tool} visualizes model attention to text prompts and the influence of each prompt token on the image. Thus, the attention visualization from {\tool} is designed to help users refine prompt tokens based on model attention. Furthermore, {\tool}'s interface has gone through multiple rounds of careful design. For instance, users can simply hover over a token to see its influence on the image in {\tool}. However, SD Web UI requires users to type keywords in order to visualize their heatmaps over the generated image. When adjusting model attention, users can drag a slider in PromptCharm to adjust attention. By contrast, they have to manually add a bracket behind a token and enter a decimal value to adjust the attention of this token when using SD Web UI. Finally, {\tool} and its features' effectiveness are confirmed through two user studies with 24 participants.}

% {

% \begin{table}[t]
%     \responseref{}
%     \centering
%     \begin{tabular}{c| c c c | c }
%          \toprule
%          Features & Promptify~\cite{brade2023promptify} & PromptMagician~\cite{feng2023promptmagician} & SD Web UI~\cite{sdwebui} & {\tool} \\
%          \midrule
%          Prompt Suggestion & \ding{51} & \ding{51} & \ding{55} & \ding{51} \\
%          Version Control
%          \bottomrule
%     \end{tabular}
%     \caption{Caption}
%     \label{tab:rel_work}
% \end{table}
% }
% }

\subsection{Interactive Support for Generative Design}
\label{subsec:creative_design}

% zaman2015gem marks2023design matejka2018dream evirgen2022ganzilla evirgen2023ganravel chilton2021visifit yan2022flatmagic chen2018forte ko2023large liu20233dall

Our work is also related to generative design. Early attempts on generative design covering both 2D~\cite{chen2018forte, zaman2015gem} and 3D design~\cite{chen2018forte, marks2023design, kazi2017dreamsketch, yumer2015semantic, chaudhuri2013attribit}, which mostly focus on assisting users in exploring a diverse set of design alternatives. For instance, Matejka et al. proposed DreamLens, an interactive system for exploring and visualizing large-scale generative design datasets~\cite{matejka2018dream}. 
%\todo{The following two papers seem less relate to interactive support.}
The most related generative design work to us includes those using AI models to support user's design~\cite{chilton2021visifit, yan2022flatmagic, liu20233dall, liu2022opal, evirgen2023ganravel, ko2023large, wang2023popblends}. 
% Chilton et al. presented VisiFit, a computational design system that supports visual blends by utilizing deep learning models~\cite{chilton2021visifit}. PopBlends extends this by leveraging a large language model to improve conceptual blending~\cite{wang2023popblends}. 
% To help digital comic professionals improve flat colorization, Yan et al. proposed FlatMagic, an interactive Photoshop plugin based on convolutional neural networks (CNNs)~\cite{yan2022flatmagic}. % \todo{I don't quite get the following sentence. "solve specific tasks in their generative design process" is unclear and sounds not relevant to our problem.}While these works only use AI models to solve specific tasks in their generative design process, a few recent works that directly use generative models for design have been proposed. \todo{discussion from here sounds more relevant to interactive support, which HCI reviewers are more interested to read.}
Evirgen et al.~proposed GANzilla, a tool that allows users to discover image manipulation directions in Generative Adversarial Networks (GANs)~\cite{evirgen2022ganzilla}. Its follow-up work, GANravel, focuses on disentangling editing directions in GANs~\cite{evirgen2023ganravel}. \responseline{To achieve this, both GANzilla and GANravel adjust the coefficients in GAN's latent space. Different from them, {\tool} supports controlling the editing effects by adjusting the model's attention to text prompt.} To generate images for news illustration, Liu et al.~proposed Opal, an interactive system based on GPT 3 (a large language model) and VQGAN~\cite{liu2022opal}. \responseline{Opal supports image style suggestion by leveraging a \textit{Sentence-BERT} for asymmetric semantic search. By contrast, {\tool} not only leverages a state-of-the-art reinforcement learning-based model, \textit{Promptist}~\cite{hao2022optimizing}, to automatically refine prompts, but also provides a database for user's exploration.} 3DALL-E integrates OpenAI's DALL-E (a text-to-image generative model), GPT-3 and CLIP to inspire professional CAD designers' 3D design work through 2D images' prototyping~\cite{liu20233dall}. As a complementary study to the prior research, Ko et al.~conducted an interview study with 28 visual artists to help the research community understand the potential and design guidelines of using large-scale text-to-image models for visual art creations~\cite{ko2023large}. Overall, our work contributes to this area through a mixed-initiative system that helps users iteratively improve text-to-image creation through a set of multi-modal prompting supports, including automated prompt refinement, modifier exploration, model attention adjustment, and image inpainting.

\subsection{Human-AI Collaboration}
\label{subsec:human-ai-collaboration}

% \todo{Since you are submitting to the blended interaction subcommittee, your reviewers may expect you to write something about Human-AI collaboration. Essentially, the prompt engineering process can be viewed as a human-AI collaboration process. Check Section 2.2 Human-AI Collaboration in Creative Work in Toby's paper http://toby.li/files/uist23-zhang-visar.pdf Note that Toby is on the PC of this subcommittee. He cannot review our work since I recently published a paper with him. But our work is likely to be sent to people like Andrew Head and Tovi Grossman, who have similar mindsets like Toby.}

%Human-AI collaboration has been an important topic in HCI research since the emergence of employing AI applications across different domains. 
Prompt engineering is a typical form of human-AI collaboration, where prompting serves as the interface between human users and generative models. As a result, the design of {\tool} is highly motivated by the recent guidelines and design principles for human-AI collaboration~\cite{amershi2019guidelines, wang2019human, liao2020questioning, dudley2018review, cai2019hello, cai2019human}. 
% \responseline{
For instance, Amershi et al.~derived 18 design guidelines about human-AI interaction based on over 150 AI-related design recommendations collected from academic and industry sources. The design of {\tool} follows two specific guidelines. First, to \textit{make clear why the system did what it did}, {\tool} provides model explanations of the AI-generated images through model attention visualization. Second, to \textit{enable the user to provide feedback during interaction with the AI system}, {\tool} enriches the feedback loop of text-to-image generation through the model attention adjustment and image inpainting. Liao et al.~ investigated user needs about XAI through interviewing 20 UX and design practitioners~\cite{liao2020questioning}. They categorized the user needs about XAI into four categories: \textit{explain the model}, \textit{explain a prediction}, \textit{inspect counterfactual}, and \textit{example based}. The design of {\tool} carefully addresses the need of \textit{explaining a prediction}. Specifically, {\tool} visualizes the model's attention to different words in the prompts by using different background colors. Moreover, when users hover over a specific word, {\tool} further highlights the corresponding parts in the generated image. 
% }
Overall, the design of {\tool} addresses two critical challenges in human-AI collaboration through its mixed-initiative interaction design: handling the imperfection of AI models and aligning the model's interpretation with the user's creative intent.

\section{User Needs and Design Rationale}
\label{sec:design}

\subsection{User Needs in Prompt Engineering and Creative Design}

% \todo{Consider to rephrase this sentence. It reads a bit complex.}
To understand the needs of users, we conducted a literature review of previous work that has done a formative study or a user study about prompt engineering~\cite{strobelt2022interactive, liu2022design, zamfirescu2023johnny, wu2022ai, ko2023large, wang2023reprompt, brade2023promptify}~\footnote{\responseref{}Note that we only consider insights from prompting with LLMs that are generalizable to text-to-image generation.} or generative design~\cite{matejka2018dream, chilton2021visifit, evirgen2022ganzilla, yan2022flatmagic}. We also reviewed the previous work that has discussed the challenges and design guidelines with generative AI~\cite{oppenlaender2022prompt, weisz2023toward}. Based on this review, we summarize five major user needs for interactive prompt engineering in text-to-image generation. 

\vspace{1mm}
\noindent \textbf{N1: Automatically recommending and revising text prompts.} Recent studies have shown that novices often struggle with writing prompts and wish to have some suggestions on how to revise their prompts~\cite{zamfirescu2023johnny, ko2023large}. For instance, through a user study with ten non-experts, Zamfirescue-Pereira et al.~found that half of the participants did not know where to start when writing text prompts to solve a given task~\cite{zamfirescu2023johnny}. After an interview with 28 visual artists who wish to use generative AI for their own work, Ko et al.~suggested that a text prompt engineering tool that can recommend or revise the prompts is an urgent need~\cite{ko2023large}. 
% \todo{I don't think the following two sentences are related to our design since we didn't follow any templates. So I would just comment them out to avoid raising questions about why we did not follow any templates.}Researchers also found that good text prompts are usually formed by specific templates~\cite{liu2022design, oppenlaender2022prompt, wang2023reprompt}. Therefore, the automatically recommended or revised prompts should follow these templates.

\vspace{1mm}
\noindent \textbf{N2: Balancing automation and user's control.} Users who use AI for creative design prefer to gain control to some extent instead of having full automation~\cite{chilton2021visifit, yan2022flatmagic, oppenlaender2022prompt, weisz2023toward}. Through a user study with five professionals in digital comics creation, 
%\todo{place tilt (~) after all et al.}
Yan et al.~emphasized the importance of automation-control balance in Human-AI co-creation~\cite{yan2022flatmagic}. Specifically, through a user study with twelve participants, Evirgen et al.~found that users wish to decide how strongly editing effects to be applied to an image generated by Generative Adversarial Networks (GANs)~\cite{evirgen2022ganzilla}. Specifically, the user appreciates changing such strength both positively and negatively.
% change the strength of some features of their created contents \todo{what does positive or negative mean here? Try to rewrite this sentence in a more clear language.}both positively and negatively when using Generative Adversarial Networks (GANs)~\cite{evirgen2022ganzilla}.

% \vspace{1mm}
% \noindent \textbf{N3: Splitting a complex task into multiple iterations with feedback.} Using generative AI to solve a complex task could be challenging for end-users. Therefore, researchers found that it is more desirable to split the task into multiple iterations~\cite{chilton2021visifit, yan2022flatmagic, wu2022ai, brade2023promptify}. Through a formative study, Chilton et al. found that one crucial design principle in an iterative design process is to structure the problem into sub-tasks and provide interactive support for each of them~\cite{chilton2021visifit}. By closely working with ML practitioners, Strobelt et al. found that a prompt engineering tool should provide the user with the human-in-the-loop ability with rich feedback through iteratively improving prompt writing~\cite{strobelt2022interactive}.

\vspace{1mm}
\noindent \textbf{N3: Supporting users explore different prompting options.} Different prompting options may yield completely different results from a generative AI model.
%\todo{"closely working with ML practitioners" sounds a bit informal. Is it a contextual inquiry study?}
Through a user-centered design process with NLP researchers, Strobelt et al.~found that a prompt engineering tool should provide the user with the human-in-the-loop ability to explore and select different variations of prompts~\cite{strobelt2022interactive}. For text-to-image creation, user needs particularly lie in the selection of image modifiers that have significant impacts on the quality and style of the generated images
%\todo{It is unclear what image modifiers refer to here. It's better to give some definition and also show some examples.} 
as mentioned in the previous studies~\cite{liu2022design, oppenlaender2022prompt, brade2023promptify}. Through a formative study with six experienced stable diffusion users, Brade et al.~highlighted the challenge of discovering effective prompting modifiers~\cite{brade2023promptify}.
% \todo{"the limited discoverability" is a bit unclear. Rewrite this sentence and provide a bit more elaboration here.}the limited discoverability of effective prompting modifiers~\cite{brade2023promptify}.
They found that users need significant effort to find keywords related to a specific image style when seeking guidance from online communities.
Therefore, {\tool} should assist users in discovering diverse image modifiers.

\vspace{1mm}
\noindent \textbf{N4: Version control to keep track through iterations.} During an iterative creating process, users may wish to compare their current contents with previous iterations at some point~\cite{evirgen2022ganzilla, weisz2023toward, brade2023promptify}. For instance, Weisz et al.~found that versioning and visualizing differences between different outputs could be helpful since users may prefer earlier outputs to later ones~\cite{weisz2023toward}. Evirgen et al.~found that keeping track of all steps could provide users more guidance once they were stuck with generative AI~\cite{evirgen2022ganzilla}.

\vspace{1mm}
\noindent\textbf{N5: Providing explanations for generated contents.} Explanations could help users better understand the generated content and gain insights for further improvements~\cite{evirgen2022ganzilla, zamfirescu2023johnny, weisz2023toward, brade2023promptify}. Zamfirescu-Pereira et al. found that users could face understanding barriers, e.g., why the model did not produce expected outputs~\cite{zamfirescu2023johnny}. Strobelt et al.~found that a prompt engineering tool should provide the user with the human-in-the-loop ability with rich feedback to iteratively improve their prompt writing~\cite{strobelt2022interactive}.
% Additionally, users could further struggle with checking these failures. 
Therefore, providing proper explanations can help calibrate users' trust with generative AI's capabilities and limitations~\cite{weisz2023toward}. Specifically, Evirgen et al.~highlighted the possibility of providing users with explanations through an informative visualization design, e.g., heat map~\cite{evirgen2022ganzilla}. 

\begin{figure*}[t]
    \centering
    \includegraphics[width=\linewidth]{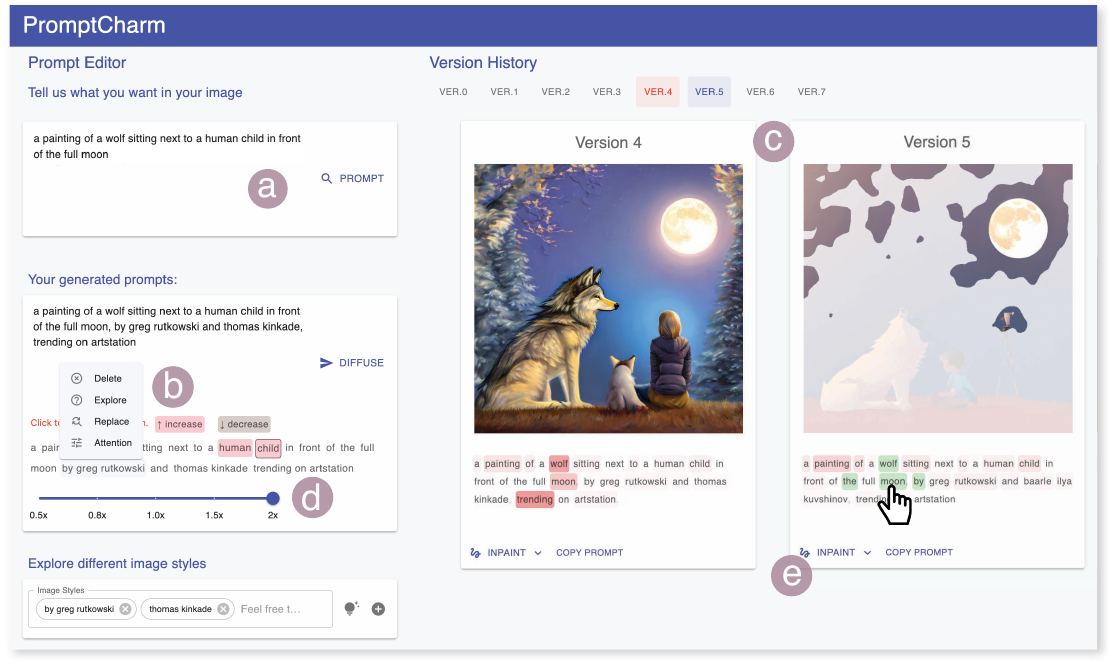}
    \caption{{\bf User interface of {\tool}.} (a) The user can first type their initial input prompt in a text box. {\tool} will then automatically refine the user's prompt. (b) {\tool} further supports users in efficiently exploring different image styles. (c) The user can then examine the generated images with the help of model attention visualization. If they would like to refine the image, they can further (d) adjust the model's attention to a keyword or (e) directly inpaint the image.}
    \Description{This figure shows the user interface of PromptCharm, which consists of two main blocks. The left block includes three different text boxes. In the top text box, the user can type their initial prompt. The machine-refined prompt will show in the middle text box, where the user can further edit and refine it. In the bottom text box, the user can type a few keywords and retrieve example images from a large database. The right block consists of a version control bar on the top and two generated images at the bottom. Each of the generated images is displayed with the raw image and model explanations by coloring each token in the prompt with different background colors.}
    \label{fig:interface}
\end{figure*}

\subsection{Design Rationale}

To support \textbf{N1}, {\tool} leverages a state-of-the-art model, \textit{Promptist}~\cite{hao2022optimizing}, to automatically revise user's input text prompt. After the user types their initial prompt and clicks on the \rcolorbox{primary}{\fontfamily{lmss}\selectfont\footnotesize \textcolor{primary}{PROMPT}} button, {\tool} will improve the user's prompt through re-organizing and appending suggested modifiers (Fig.~\ref{fig:interface}~\circled{a}). In order to balance automation and user's control (\textbf{N2}), {\tool} provides multi-modal prompting within an interactive text box (Fig.~\ref{fig:interface}~\circled{b}). In this text box, the user can \textbf{Delete}, \textbf{Explore}, or \textbf{Replace} one or multiple image modifiers. By exploring different modifiers, the user can know how these keywords would possibly affect the generated images without actually rendering it. If the user encounters any modifiers that they dislike, they can efficiently replace them with other similar/dissimilar modifiers in {\tool}. This design also allows the user to explore different prompting options (\textbf{N3}). 
Note that an alternative design choice for \textbf{N3} could be generating and displaying a large set of images for the user to explore given different suggested prompts. However, such a design might be overwhelming for novice users. \responseline{For example, when leveraging text-to-image models for news illustration, Liu et al. found that returning a large amount of choices could at times be \textit{overwhelming, repetitive, or over specifc}~\cite{liu2022opal}. Moreover, during our experiments, we found that generating one image with a NVIDIA A5000 GPU (24 GB VRAM) could take around 30 seconds. Generating a large set of images would require much more computational resources or introduce a longer waiting time for users.} Thus, {\tool} focuses on helping users iteratively refine their images instead of rendering a set of images for users to select.
% \todo{It would be nice to discuss some alternative design choices here, like rendering many generated images in a cluster view. Explain why a cluster view is not a good choice here and how your design is better than a cluster view. I have recently received several reviews that explicitly asked me to discuss alternative designs.}

To guide users iteratively improving their creations, {\tool} provides model explanations by visualizing the model's attention after each iteration (\textbf{N5}). By observing the attention values of different words in the input prompt, the user can quickly identify if the model's interpretation of the prompt was aligned with their intent, e.g., whether there is an important keyword the model does not pay attention to (Fig.~\ref{fig:interface}~\circled{c}). Moreover, {\tool} assists users in interpreting the model's generation by highlighting the parts of the image that correspond to a keyword in the text prompt.
%\todo{" through a heat-map visualization when hovering over a specific word in the prompt" is unclear. Please rewrite it. Maybe break it into two sentences.}through a heat-map visualization when hovering over a specific word in the prompt. 
% \todo{The following sentence is too long. Simplify it.}
If the user observes any misalignment between the model's interpretation and their creative intent, they can refine the generated image through adjusting the model's \textbf{Attention} to the keywords in the prompts (Fig.~\ref{fig:interface}~\circled{d}). If the generated image includes undesired parts, the user can also mask these parts and re-generate them through an image inpainting model in {\tool} without spending more effort on rewriting prompts (Fig.~\ref{fig:interface}~\circled{e}). This is highly motivated by the idea of direct manipulation~\cite{shneiderman1981direct, shneiderman1982future}. Both the model attention adjustment and the image inpainting can also support \textbf{N2}. Finally, to support \textbf{N4}, {\tool} provides version control to help users keep track through multiple iterations. By clicking on the labels of different versions, the user can quickly glance and compare their text prompts and synthesized images from different rounds of iterations. 
% \responseline{Note that, an alternative design choice for \textbf{N4} would be clustering and displaying all generated images in one browser~\cite{brade2023promptify, feng2023promptmagician}. {\tool} focuses on supporting users directly compare \textit{version-to-version} generated images and model explanations and avoids potential overwhelm for novice users as much as possible. Therefore, we did not use a cluster view to display the user's generated images.}

\section{Design and Implementation}
\label{sec:approach}

In this section, we introduce the design and implementation of {\tool}. Specifically, we first introduce {\tool}'s {\em multi-modal} prompting and refinement: (1) Text Prompt Refinement, Suggestion, and Exploration (Sec.~\ref{subsec:prompter}). (2) Model Attention-based Explanation and Refinement (Sec.~\ref{subsec:attention_adjuster}). (3) Direct Manipulation via Inpainting and Masked Image Generation (Sec.~\ref{subsec:inpainting}).
% \todo{The organization here is different from the ordering in abstract and intro. It's better to organize it as (1) "Textual Prompt Refinement, Suggestion, and Exploration". (2) Attention-based Explanation and Refinement. (3) Direct Manipulation via Inpainting and Masked Image Generation.}(1) prompt editing and refinement (Sec.~\ref{subsec:prompter}), (2) image refinement via attention adjustment (Sec.~\ref{subsec:attention_adjuster}), and (3) image refinement via inpainting (Sec.~\ref{subsec:inpainting}). 
Then we introduce the iterative creation process with version control in {\tool} (Sec.~\ref{subsec:version_control}).

\subsection{Text Prompt Refinement, Suggestion, and Exploration}
\label{subsec:prompter}

\begin{figure*}[t]
    \centering
    \includegraphics[width=\linewidth]{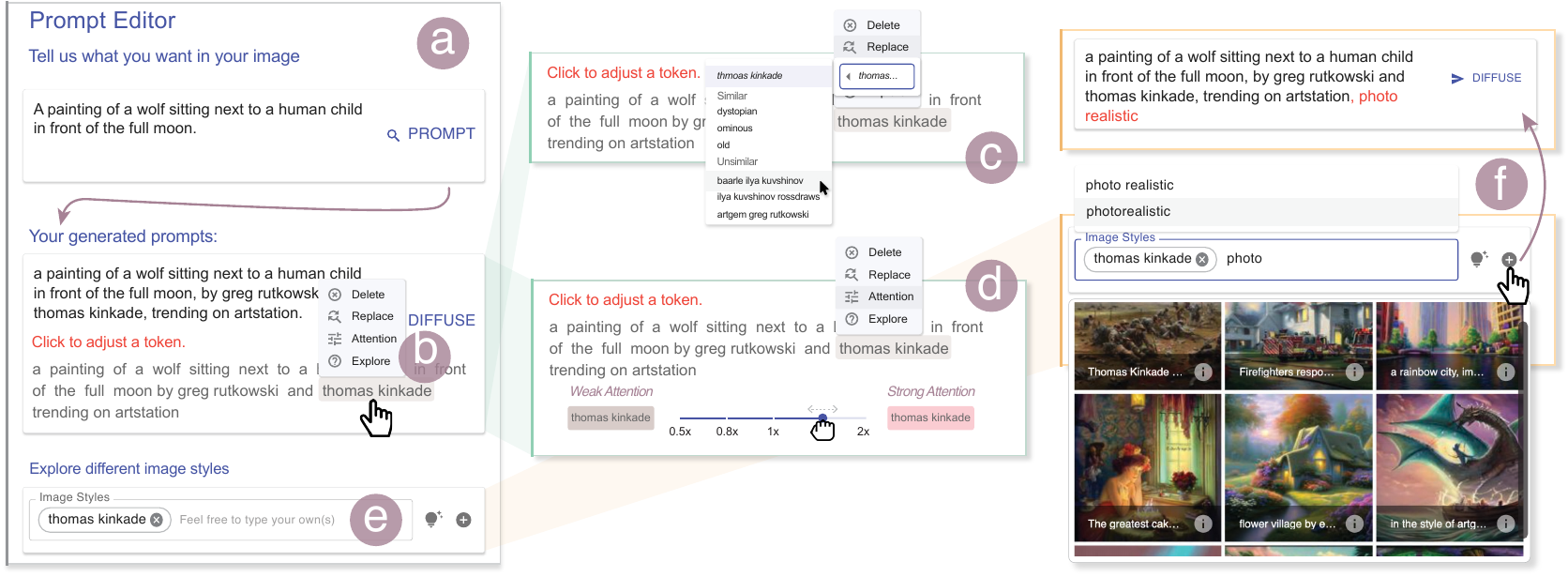}
    \caption{{\tool} provides (a) automated prompt refinement and (b) prompt editing in the \textit{prompting view}. The user can (c) replace a modifier with similar/dissimilar styles, (d) adjust the model's attention to a keyword, or (e) explore popular modifiers and (f) append them to the prompt.}
    \Description{This figure displays the interaction of PromptCharm’s text prompt refinement, suggestion and exploration. The top-left corner (a) of this figure includes a text box, where the user can type a few keywords as their initial prompts. Right behind this text box is another text box showing the machine-refined prompts (b). The user can click any of the keywords in this refined prompt, and a menu with four options: delete, replace, attention, and explore will pop up. When the user clicks on replace (c), they can replace the selected keyword with similar/dissimilar image styles in a drop-down list. When the user clicks on attention (d), they can drag a slider to control the model’s attention to this keyword. In the bottom text box, the user can type a few keywords or select from the prompt (e). PromptCharm will display the retrieved images from a large database. The user can further click on an “add” icon to append the explored keywords to their prompt (f).}
    \label{fig:prompting_view}
\end{figure*}

\smalltitle{Automated prompt refinement.} To help users refine their initial input prompts, {\tool} leverages a state-of-the-art prompt optimization model released by Microsoft, \textit{Promptist}~\cite{hao2022optimizing}. Promptist is designed to rephrase users' input prompts while retaining their original intentions. It is based on the GPT-2 architecture and was initially trained using a dataset that consisted of pairs of user input prompts and prompts refined by engineers. After supervised training, Promptist was then fine-tuned through reinforcement learning to optimize the prompt for generating visually pleasing images for \textit{Stable Diffusion}. {\tool} uses a pre-trained Promptist released by the authors~\footnote{\href{https://huggingface.co/microsoft/Promptist}{https://huggingface.co/microsoft/Promptist}}.

In {\tool}, the user can type their own words to describe what they would like to see in their generated images in a text box (Fig.~\ref{fig:prompting_view}~\circled{a}). When the user click on the button \rcolorbox{primary}{\fontfamily{lmss}\selectfont\footnotesize \textcolor{primary}{PROMPT}}, {\tool} will refine this input prompt with Promptist (Fig.~\ref{fig:prompting_view}~\circled{a}). The refined prompt will show up in the middle text box for the user to compare with the initial prompt. When the user clicks on the button \rcolorbox{primary}{\fontfamily{lmss}\selectfont\footnotesize \textcolor{primary}{DIFFUSE}}, the diffusion model will generate a new image (Fig.~\ref{fig:prompting_view}~\circled{b}).

\smalltitle{Suggesting popular modifiers.} Previous studies have discussed the importance of \textit{modifiers} (keywords that have significant effects on the generated image's style and quality) and their impacts~\cite{liu2022design, oppenlaender2022prompt}. To assist users in efficiently exploring different modifiers and image styles, {\tool} leverages a data-mining method to extract popular modifiers from a dataset of text-to-image prompts, \textit{DiffusionDB}~\cite{wang2022diffusiondb}. {\tool} first applies the CountVectorizer algorithm~\cite{pedregosa2011scikit} to extract top frequent modifiers from the \textit{DiffisuionDB}. Given that a modifier may consist of several words (tokens), we consider $n$-gram phrases during our mining process, where $n=1, 2, 3$. Finally, {\tool} removes modifiers that only include ``stop words'' and sorts them according to the frequency. 

When the user clicks on a token in the refined prompt in {\tool}, a menu with four different options (\textbf{Delete}, \textbf{Replace}, \textbf{Attention}, and \textbf{Explore}) will pop up (Fig.~\ref{fig:prompting_view}~\circled{b}). When the user clicks on \textbf{Replace}, \responseline{the top-3 image modifiers mined from DiffusionDB~\cite{wang2022diffusiondb} that provide the most similar and dissimilar art effects} will show up in a drop-down list. The user can then choose to replace the selected modifier with one of them (Fig.~\ref{fig:prompting_view}~\circled{c}). Note that the similarity between two image modifiers is calculated based on the cosine distance between their embedding obtained from the diffusion model's \textit{text encoder}. When the user clicks on \textbf{Attention}, a slider will pop up for the user to adjust the model's attention to the selected keyword(s) (Fig.~\ref{fig:prompting_view}~\circled{d}). We will introduce the details of the model's attention adjustment in Sec.~\ref{subsec:attention_adjuster}. When the user clicks on \textbf{Explore}, the selected keyword(s) will be added to the bottom text field for further exploration (Fig.~\ref{fig:prompting_view}~\circled{e}). 

\smalltitle{Image Style Exploration.} In addition to replacing with similar/dissimilar image styles, the user can further explore and pick their own choices in {\tool} (Fig.~\ref{fig:prompting_view}~\circled{e}). {\tool} displays popular image modifiers in a drop-down list  (Fig.~\ref{fig:prompting_view}~\circled{f}). The user can select from these popular modifiers, type a few new modifiers, or select from the prompt. After entering a few modifiers, the user can click on the ``bulb'' {\promptsearchicon\,} icon to search for images that include these modifiers from  DiffusionDB~\cite{wang2022diffusiondb}. The search results will be displayed in a pop-up grid. The user can then hover over a specific image of interest to check its complete text prompt. If the user is satisfied with the selected modifiers, they can click on the ``add'' {\promptaddicon\,} icon. These modifiers will be appended to the prompt in the middle text box (Fig,~\ref{fig:prompting_view}~\circled{f}).

\subsection{Model Attention-based Explanation and Refinement}
\label{subsec:attention_adjuster}

\begin{figure*}[t]
    \centering
    \includegraphics[width=\linewidth]{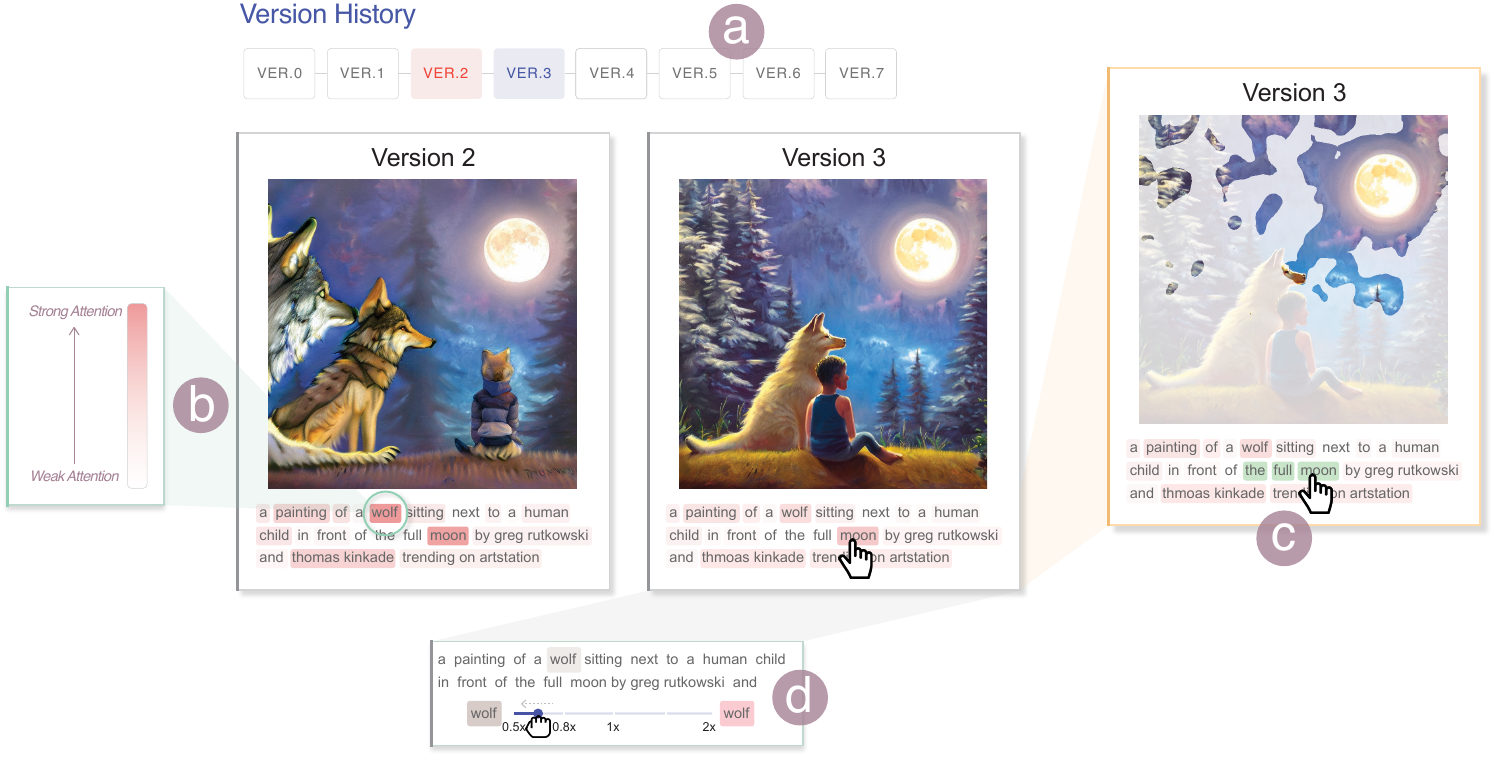}
    \caption{{\tool} provides version control with XAI feedback to help users iteratively improve their creations. (a) The user can efficiently switch between different versions. (b) The user can also observe the model's attention to each token. (c) The user can further hover over a token to check the corresponding parts in the generated image. (d) Once the user notices any ``over-attending'', they can directly adjust the attention to specific keywords. For example, by adjusting the attention of the word, ``wolf'', the model avoids mis-attending to the ``human child'' during the generation.}
    \Description{This figure displays PromptCharm’s model attention visualization and attention adjustment. The top of this figure includes a version control bar. The user can switch between different versions by clicking on different versions’ labels (a). The middle of this figure includes two different images, where the left one includes three wolves and the right one includes a wolf and a human child. The model’s attention to different keywords is visualized by colouring each keyword with different background colors at the bottom of each image. The darker the color, the higher the attention value (b). When the user hovers over a keyword, PromptCharm will also highlight the corresponding parts of the generated images (c). If the user would like to change a keyword’s attention, they can drag the slider to control it (e).}
    \label{fig:version_control}
\end{figure*}

\smalltitle{Model attention visualization.} The attention mechanism is the most important component in the transformer models for the purpose of capturing semantic relationships among different tokens/pixels. The Stable Diffusion model, as a multi-modal model, leverages a \textit{cross-attention} mechanism to unify the model's attention on the input text prompt and the generated image. The cross-attention scores can further be used to interpret the model's generation~\cite{tang2022daam}. Therefore, considering the user might be curious about how a generated image is correlated to their input prompts~\cite{weisz2023toward, evirgen2022ganzilla}, {\tool} renders model explanations over both the input text prompt and the generated image with a attention-based XAI technique, \textit{DAAM}~\cite{tang2022daam}. DAAM can generate heat-map explanations over a stable diffusion model's generated image. Given a specific word from the input prompt, DAAM aggregates the model's cross-attention scores across layers and projects it to the generated image. The attention score of each token is further used to represent a token's saliency in {\tool} to help users understand its importance towards the generation. 

{\tool} visualizes the model's attention in two ways: First, each token in the prompt is colored according to its saliency (Fig.~\ref{fig:version_control}). The higher importance a token has contributed to the generation, the darker background color it is assigned (Fig.~\ref{fig:version_control}~\circled{b}). Second, when the user hovers over a token, {\tool} will highlight particular parts of the image that are strongly related to this token during generation (Fig.~\ref{fig:version_control}~\circled{c}). 

Finally, {\tool} can also help users interpret the correlations among different tokens in the text prompt. Given a selected token from the prompt, {\tool} leverages a neuron activation analysis to extract a set of tokens that have similar contributions to the generated images~\cite{alammar2021ecco}. When the user hovers over a token in {\tool} (Fig.~\ref{fig:version_control}~\circled{c}), the corresponding set of similar tokens will be highlighted with a different background color.

% \todo{Not sure why starting from model attention misalignment. It's better to talk about the attention explanation method (Section 4.4) and then talk about misalignment and then talk about the adjustment as a feedback.}

\smalltitle{Model attention misalignment.} Given a prompt, a model's attention is normally automatically calculated. However, previous studies have shown that a transformer model's calculated attention could be massively misaligned with the user's intention~\cite{kou2023model, garcia2023uncurated}. Specifically, for text-to-image generation, the generated image's content could be inconsistent with the text prompt. Fig.~\ref{fig:version_control}~\circled{d} shows an example of such attention misalignment. In this example, the original image (left one) misses the object ``human child'' in the text prompt while rendering an extra object of ``wolf''. To address this, {\tool} uses an interactive design that allows users to adjust the model's attention to keywords in the prompt through a slider (Fig.~\ref{fig:version_control}~\circled{d}). Such design is highly inspired by the recent studies on aligning human's attention with the transformer model's attention~\cite{wang2022kvt, wang2022matchformer, he2023efficient}. As depicted in Fig.~\ref{fig:version_control}~\circled{d}, by decreasing the model's attention to the ``wolf'', the model then correctly generate an image with one ``wolf'' and one ``human child.''

% \begin{figure}[h]
%     \centering
%     \includegraphics[width=0.7\textwidth]{attention.pdf}
%     \caption{By adjusting the attention of the word, ``wolf'', the model avoids mis-attending to the ``human child'' during the generation.}
%     \label{fig:attention}
% \end{figure}

\smalltitle{Model attention adjustment.} To adjust the model's attention, {\tool} utilizes the \textit{hooking} technique. The hooking technique originates from the software instrumentation, which allows a developer to do run-time modification over a running software process. {\tool} places hooks on all $l$ \textit{cross-attention layers} of a given stable diffusion model without modifying its architecture. Algorithm~\ref{alg:attention} further depicts {\tool}'s model attention adjustment process. Suppose the input prompt is $\mathbf{x}=[x_1, x_2, \dots, x_n]$, where $x_j$ denotes the $j$th token. Given a set of user-selected tokens $\Tilde{\mathbf{x}}$ subject to attention adjustment with corresponding adjustment factors $\mathbf{\gamma}$, the algorithm first extracts the unaltered output $\mathcal{S}^i$ at the $i$th cross-attention layer (Line 3). Then, for each token $x_j$ that is subject to attention adjustment, {\tool} multiplies the cross-attention output at $j$th token $\mathcal{S}^i_j$ with the user-defined factor $\mathbf{\gamma}_j$ (Line 5:10). The new text feature $\mathcal{F}^i$ is then calculated based on the altered cross-attention output (Line 11). This process loops until the model $\mathbf{M}$ finishes all $l$ layers' inferences. During the user study, we set the maximum and minimum values of $\gamma_j$ as 2 and 0.5 to avoid over-attending or completely miss-attending.

\begin{algorithm}[t]
\caption{The algorithm of adjusting a diffusion model's attention.}
\label{alg:attention}
\KwIn{the stable diffusion model $SD$ with $l$ cross-attention layers, the input text prompt $\mathbf{x}$ of $n$ tokens, the list of tokens subject to adjustment $\Tilde{\mathbf{x}}$, the list of attention adjustment factors $\mathbf{\gamma}$}
\KwOut{the adjusted model attention $\mathcal{S}^l$}
$\mathcal{F}^0~\gets~\mathsf{intial~feature}$\;
\For{$i~$in$~1,\dots,l$}{
    $\mathcal{S}^i~\gets~\mathsf{cross\_attention}(SD, \mathcal{F}^{i-1})$\;
    \For{$j~$in$~1,\dots,n$}{
        \eIf{$x_j~$in$~\Tilde{\mathbf{x}}$}{
                $\mathcal{S}^i_j~\gets~\mathcal{S}^i_j\times\mathbf{\gamma}_j$\;
            }{
                $\mathcal{S}^i_j~\gets~\mathcal{S}^i_j$\;
            }
    }
    $\mathcal{F}^i~\gets~\mathsf{FFN}(\mathcal{F}^{i-1}, \mathcal{S}^i)$\;
}
\KwRet{$\mathcal{S}^l$}\;
\end{algorithm}

\begin{figure*}[t]
    \centering
    \includegraphics[width=\linewidth]{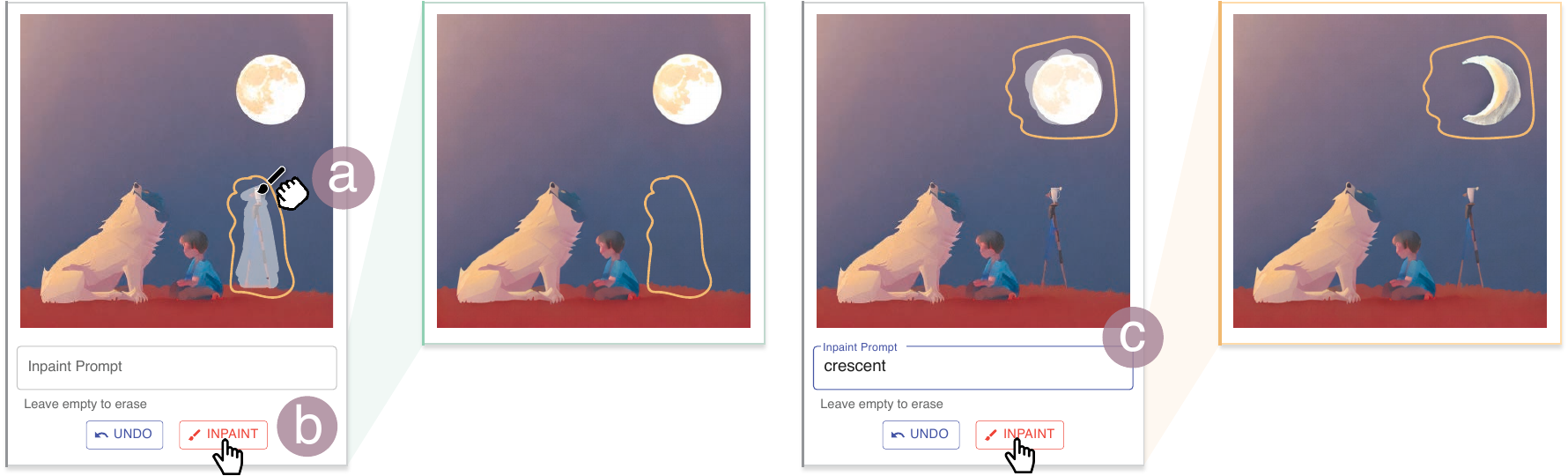}
    \caption{The user can (a) mask the undesired areas in a generated image then (b) re-generate these areas in {\tool}. (c) The user can further provide text prompts to guide the inpainting process.}
    \Description{This figure shows the inpainting feature in PromptCharm through four different images from left to right. In the first image, the user brushes over an area of a ladder and clicks on the inpaint button. The inpainting result is shown in the second image, where the ladder is erased. In the third image, the user brushes over the moon and types the prompt “crescent” in the text box. Once the user clicks on the inpaint button, the fourth image shows the inpainting result, where the full moon is replaced by a crescent.}
    \label{fig:inpainting}
\end{figure*}

\subsection{Direct Manipulation via Inpainting and Masked Image Generation}
\label{subsec:inpainting}

In addition to \textit{Attention Adjustment}, {\tool} further supports refining AI-generated images through \textit{image inpainting}. The purpose of this image inpainting feature is to allow users to re-render the small undesired areas in a generated image without further crafting the prompt. This is highly motivated by the idea of direct manipulation~\cite{shneiderman1981direct, shneiderman1982future}. To achieve this, we use the stable diffusion model with the image inpainting pipeline~\footnote{\href{https://huggingface.co/docs/diffusers/api/pipelines/stable_diffusion/inpaint}{https://huggingface.co/docs/diffusers/api/pipelines/stable\_diffusion/inpaint}}. 

To inpaint an image, the user can brush over the area of a generated image $I$ that they would like to re-generate to create a mask $M$ in {\tool} (Fig.~\ref{fig:inpainting}~\circled{a}). Given the image $I$ and a mask $M$, when the user clicks on the \rcolorbox{secondary}{\fontfamily{lmss}\selectfont\footnotesize \textcolor{secondary}{INPAINT}} button, {\tool} will render a new image $I^\prime$ where only pixels masked by $M$ would be changed (Fig.~\ref{fig:inpainting}~\circled{b}). The user can further provide a text prompt $\mathbf{x}_{inpaint}$ they would like to incorporate in guiding the inpainting process by typing in the text box (Fig.~\ref{fig:inpainting}~\circled{c}). If $\mathbf{x}_{inpaint}$ is not given, the inpainting will be solely guided by the unmasked areas of the image.

% \todo{Don't describe the view and the implementation separately. It feels strange. Interleave them. }
% \smalltitle{Inpainting view.} In this view, the user can brush over the area of a generated image that they would like to re-generate (Fig.~\ref{fig:inpainting}~\circled{a}). When the user click on the \rcolorbox{secondary}{\fontfamily{lmss}\selectfont\footnotesize \textcolor{secondary}{INPAINT}} button, {\tool} will execute the inpainting pipeline to render an inpainted image (Fig.~\ref{fig:inpainting}~\circled{b}). The user can further provide the text prompt they would like to incorporate in guiding the inpainting process through typing in a text box (Fig.~\ref{fig:inpainting}~\circled{c}).

\subsection{Iterative Creation with Version Control}
\label{subsec:version_control}

% \todo{Talk about version control as a separate feature in a separate subsection.}
% \smalltitle{Version control.} 
{\tool} integrates a version control component to assist users in keeping track on their generated images. Each version of the generated images is displayed with the corresponding model explanations (Fig.~\ref{fig:version_control}). When the user clicks on one or multiple specific versions, the corresponding generated images and model explanations will show up (Fig.~\ref{fig:version_control}~\circled{a}). The user can switch between different versions to examine their changes over different iterations. By default, {\tool} presents two versions at the same time to help users compare the prompts and generated images side-by-side.

\subsection{Implementation}
\label{subsec:implementation}

We implemented and deployed {\tool} as a web application. The user interface of {\tool} is implemented with Material UI~\footnote{\href{https://mui.com}{https://mui.com}}. The back-end of {\tool} is based on Python Flask. All machine learning models are implemented with PyTorch~\footnote{\href{https://pytorch.org}{https://pytorch.org}} and Transformers~\footnote{\href{https://github.com/huggingface/transformers}{https://github.com/huggingface/transformers}}. For the diffusion model, we use the Stable Diffusion v2-1~\footnote{\href{https://huggingface.co/stabilityai/stable-diffusion-2-1}{https://huggingface.co/stabilityai/stable-diffusion-2-1}}. During user studies, {\tool} ra on a server with two NVIDIA A5000 GPUs.

\section{Usage Scenario}
\label{sec:scenario}

\begin{figure*}[t]
     \centering
     \begin{subfigure}[b]{0.18\textwidth}
         \centering
         \includegraphics[width=\textwidth]{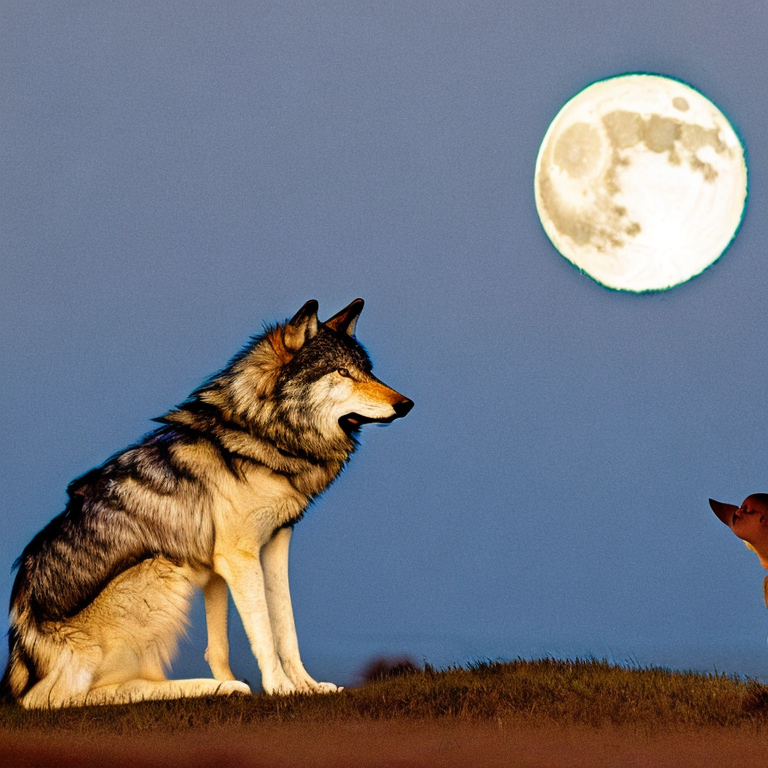}
         \caption{Iteration 1}
         \label{fig:step_1}
     \end{subfigure}
     \hfill
     \begin{subfigure}[b]{0.18\textwidth}
         \centering
         \includegraphics[width=\textwidth]{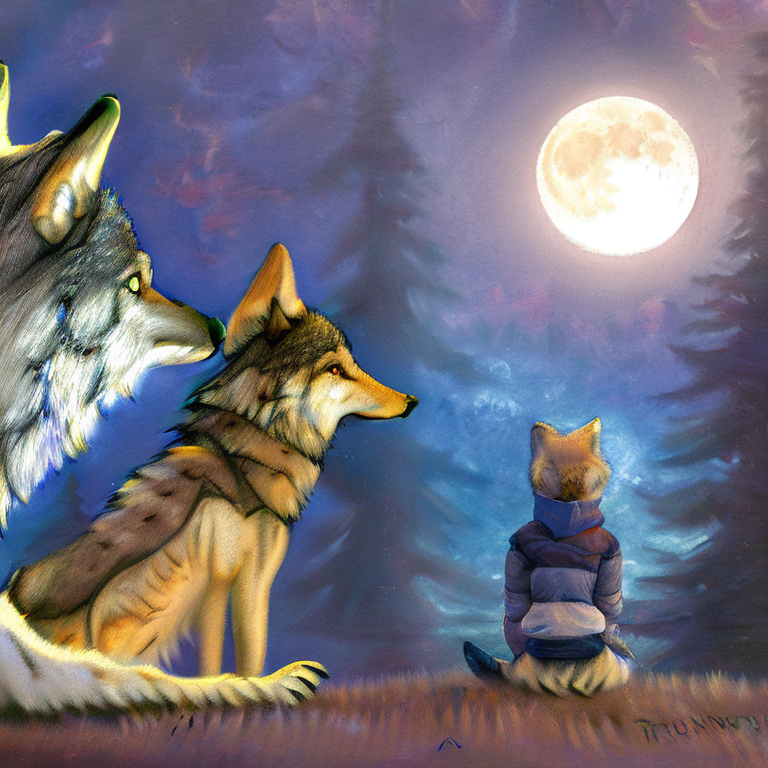}
         \caption{Iteration 2}
         \label{fig:step_2}
     \end{subfigure}
     \hfill
     \begin{subfigure}[b]{0.18\textwidth}
         \centering
         \includegraphics[width=\textwidth]{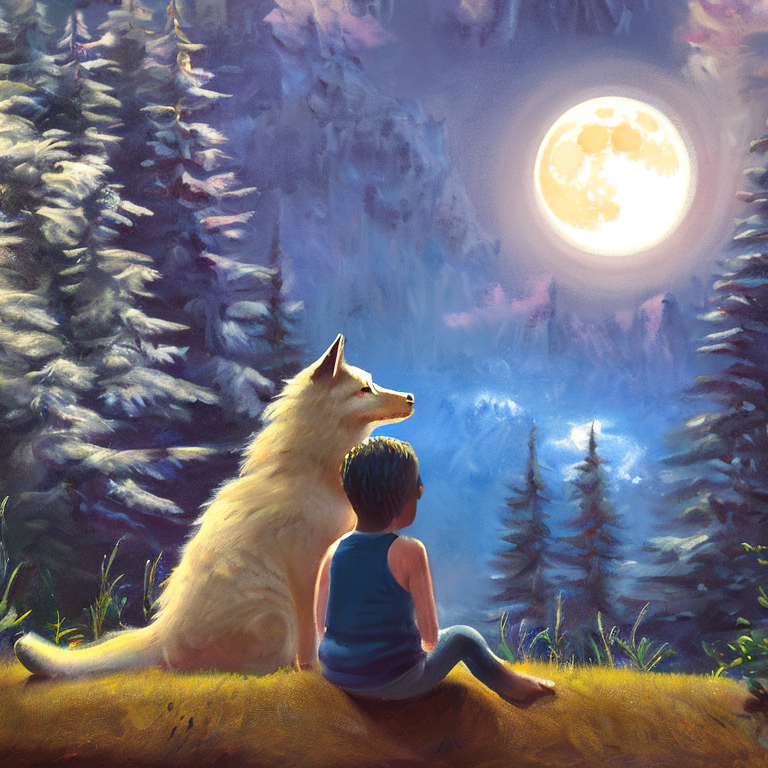}
         \caption{Iteration 3}
         \label{fig:step_3}
     \end{subfigure}
     \hfill
     \begin{subfigure}[b]{0.18\textwidth}
         \centering
         \includegraphics[width=\textwidth]{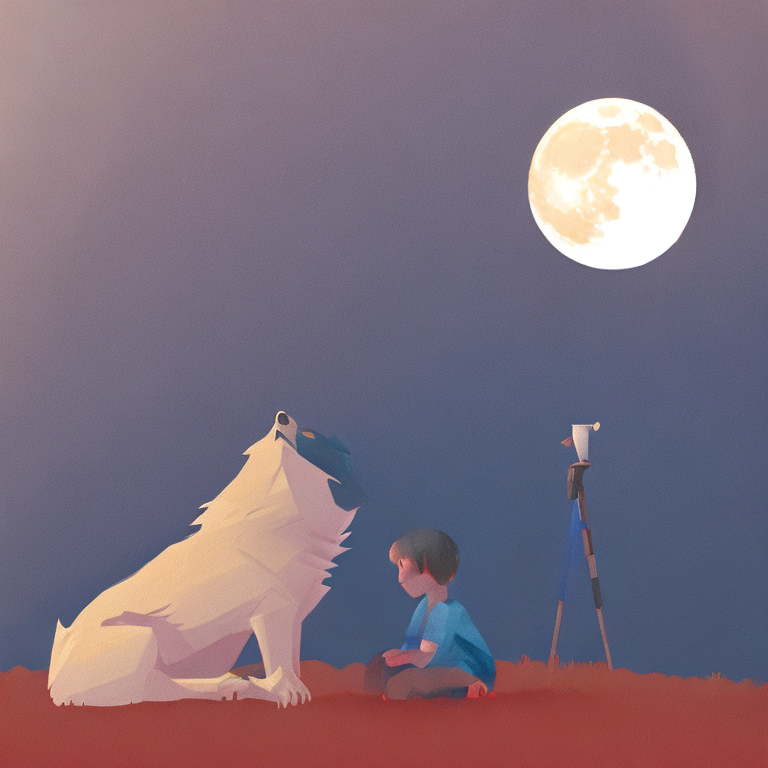}
         \caption{Iteration 4}
         \label{fig:step_4}
     \end{subfigure}
     \hfill
     \begin{subfigure}[b]{0.18\textwidth}
         \centering
         \includegraphics[width=\textwidth]{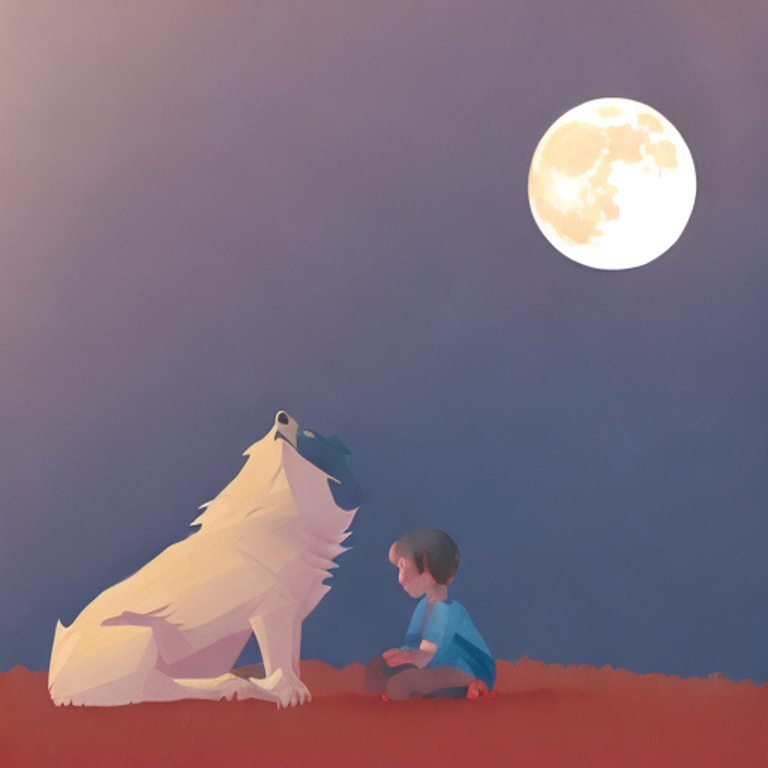}
         \caption{Iteration 5}
         \label{fig:step_5}
     \end{subfigure}
    \caption{An example scenario of iteratively improving an image creation featuring ``\textit{a wolf sitting next to a human child in front of the full moon.}''}
    \Description{This figure shows the six iterations of the user’s creations through six different images from left to right. The first image only contains a wolf. The second image includes three wolves. The third image includes a wolf and a human child. The fourth image includes a wolf and a human child while in a different image-style compared with the third image. The last image excludes the ladder in the fourth image.}
    \label{fig:usage_scenario}
\end{figure*}

Suppose Alice is a novice user who would like to use the Stable Diffusion model to create an image. The image in her mind features the following content: ``\textit{a wolf sitting next to a human child in front of the full moon.}'' However, upon entering this prompt, she discovers that the model's generated result does not align with the image she envisions. Specifically, she observes that the ``human child'' is not positioned correctly in the image (Fig.~\ref{fig:step_1}). Additionally, she finds that her generated image lacks a sense of aesthetics. Alice then searches for relevant tutorials on the Internet. In her quest for text-to-image prompt examples on the Internet, Alice encounters numerous online resources. Nevertheless, she finds it cumbersome to experiment with various examples. Therefore, Alice decides to give {\tool} a try.

Alice first checks the \textit{Prompting View}, where she can articulate her requirements within a designated text box (Fig.~\ref{fig:interface}~\circled{a}). By clicking on the button \rcolorbox{primary}{\fontfamily{lmss}\selectfont\footnotesize \textcolor{primary}{PROMPT}}, {\tool} autonomously generates a fresh prompt and presents it in another text box for her to further refine (Fig.~\ref{fig:interface}~\circled{b}).  Alice observes that this generated prompt incorporates a few additional modifiers, such as ``by greg rutkowski'', ``thomas kinkade'', and ``trending on artstation.'' She ponders how these modifications might enhance the resulting image. Subsequently, she opts to generate a new image by clicking on the button \rcolorbox{primary}{\fontfamily{lmss}\selectfont\footnotesize \textcolor{primary}{DIFFUSE}}. The new image shows up on the right side of the interface (Fig.~\ref{fig:interface}~\circled{c}) after the model finishes generation.  Alice is pleased to note that the newly generated image exhibits substantial improvement in terms of the visual effects (Fig.~\ref{fig:step_2}). 

Nevertheless, she observes that the model fails to include the ``human child'' in the image and, intriguingly, includes multiple instances of the ``wolf'' object. Therefore, Alice resolves to refine her prompt with {\tool}'s assistance to rectify her image. Alice discerns that the word ``wolf'' has a very high model attention value (Fig.~\ref{fig:version_control}~\circled{d}). When she hovers over the word ``wolf'', a large portion of the image is highlighted. She interprets the model may probably over-attend to the word ``wolf''. Therefore, she clicks on the word ``wolf'' and select the \textbf{Attention} option. She proceeds to reduce the model's attention to this word by a factor of 0.5 before regenerating the image (Fig.~\ref{fig:version_control}~\circled{d}). As a result, the image now accurately features both the ``human child'' and the ``wolf'' objects (Fig.~\ref{fig:step_3}).

Alice is now much more satisfied with the image she has created. She ponders which modifier has contributed to the enhanced visual effects. She clicks on the modifier ``thomas kinkade'' and selects the \textbf{Explore} option. In the ensuing pop-up panel (Fig.\ref{fig:prompting_view}~\circled{f}), she discovers quite a few examples from the database that exhibit similar visual effects to her generated image.  This exploration provides Alice with a rudimentary understanding of the image's style. She believes the modifier ``thomas kinkade'' has contributed to such special atmosphere of the generated image. Nevertheless, her curiosity prompts her to experiment with other diverse image styles. Therefore, she selects the \textbf{Replace} option (Fig.~\ref{fig:prompting_view}~\circled{c}). In a drop-down list, she opts to replace the modifier ``thomas kinkade'' with a dissimilar style ``baarle ilya kuvshinov''. 

After exploring this new style, Alice finds herself more inclined toward this fresh aesthetic. Alice renders another new image (Fig.~\ref{fig:step_4}). This time, the image aligns closely with her expectations, save for a minor flaw---an additional object on the right side of the image. However, she refrains from modifying her prompt to rectify this issue, as such adjustments could potentially impact other areas in the image. Alice then attends to the \textit{inpainting view} of {\tool}. She clicks on the button \rcolorbox{secondary}{\fontfamily{lmss}\selectfont\footnotesize \textcolor{secondary}{INPAINT}} to open the inpainting canvas. Alice then brushes over the area that she would like to re-generate (Fig.~\ref{fig:inpainting}~\circled{a}). Following the inpainting process, the generated image now aligns better with her expectations (Fig.~\ref{fig:step_5}). By clicking on the label ``VER.0'' (Fig.~\ref{fig:version_control}~\circled{a}), Alice compares her most recent iteration with her initial creation. She observes a significant improvement in terms of image quality and visual effects.

\section{User Study 1: Close-ended Tasks}
\label{sec:user_study_1}

To evaluate the usability and effectiveness of {\tool}, we conducted a within-subjects user study with 12 participants from different levels of experience with generative AI and text-to-image generation models. To better understand the value of proposed features in {\tool}, we compared {\tool} with two variants of {\tool} as the baselines by disabling interactive features. Specifically,

\begin{itemize}
    \item \textbf{Baseline}. The baseline includes a plain text editor for users to write their prompts as well as the version history.

    \item \textbf{Promptist}. In addition to the features in Baseline, we further introduce the \textit{Promptist} model~\cite{hao2022optimizing} to help users refine their prompts. This is to simulate the situation in which users only have a fully automated model for prompt engineering.
\end{itemize}

\begin{table*}[t]
    \newcommand{\midsepremove}{\aboverulesep = 0mm \belowrulesep = 0mm}
    \midsepremove
    \newcommand{\midsepdefault}{\aboverulesep = 0mm \belowrulesep = 0mm}
    \midsepdefault
    \renewcommand{\arraystretch}{1.1}
    \centering
    \caption{User performance in Study 1 measured by SSIM~\cite{wang2004image}. Higher SSIM indicates better performance.}
    \begin{tabular}{l cc c cc c cc c | c cc c}
    \toprule
    \multirow{2}{*}{Conditions} & \multicolumn{2}{c}{Task 1} && \multicolumn{2}{c}{Task 2} && \multicolumn{2}{c}{Task 3} &&& \multicolumn{2}{c}{Overall} & \\
    \cmidrule{2-3} \cmidrule{5-6} \cmidrule{8-9} \cmidrule{12-13}
    & Mean & SD & & Mean & SD & & Mean & SD & & & Mean & SD &\\
    \midrule
    Baseline & $0.521$ & $0.053$ && $0.473$ & $0.082$  && $0.443$ & $0.187$ &&& $0.479$ & $0.115$ &\\
    Promptist & $0.707$ & $0.079$ && $0.529$ & $0.017$ && $0.487$ & $0.043$ &&& $0.574$ & $0.111$ &\\
    \cellcolor{tab_purple}{\tool} & \cellcolor{tab_purple}$0.708$&\cellcolor{tab_purple}$0.063$ &\cellcolor{tab_purple} &\cellcolor{tab_purple}$0.606$ &\cellcolor{tab_purple}$0.078$ &\cellcolor{tab_purple} &\cellcolor{tab_purple}$0.631$ &\cellcolor{tab_purple}$0.138$ &\cellcolor{tab_purple} &\cellcolor{tab_purple} & \cellcolor{tab_purple}$0.648$ &\cellcolor{tab_purple}$0.100$ &\cellcolor{tab_purple}\\
    \bottomrule
    \end{tabular}
    \label{tab:close-end}
\end{table*}

\subsection{Methods}
\label{subsec:methods_1}

\subsubsection{Participants}
\label{subsubsec:participants_1}

We recruited 12 participants through mailing lists of the ECE and CS departments at a research university.\footnote{This human-participated study is approved by the university’s research ethics office.} 1 participant was an undergraduate student, 3 participants were Master students, 7 were Ph.D. students, and 1 were professional developers. Participants were asked to self-report their experience with general generative AI (e.g., ChatGPT) and text-to-image creation models (e.g., DALL-E and Midjourney). Regarding experience with general generative AI, 3 participants had 2-5 years of experience, 5 had 1 year, and 4 had less than 1 year or no experience. Regarding text-to-image generative model experience, 3 participants had 1 year of experience while the remaining had less than 1 year or no experience. All participants mentioned that they had never used any prompt engineering tools before. We conducted all user study sessions via Zoom. {\tool} and two baselines were all deployed as web applications. Therefore, participants were able to access our study sessions from their own PCs.

\subsubsection{Tasks}
\label{subsubsec:task_1}

The goal of this study is to identify the effectiveness of {\tool} when helping users achieve a specific image creation target. To this end, we designed three different tasks based on the prompts from \textit{DiffusionDB} that were excluded from {\tool}'s data mining process (Sec.~\ref{subsec:prompter}). Specifically, we randomly sampled 60 prompts that included animals as subjects. Then, we used the stable diffusion model to generate an image for each of the given prompts. We resolved to select three prompts that are complex, and the corresponding generated images are with good quality and visual effects. For each task, a participant was then given one of the three images as the target. The participant's goal is to replicate the given target image as similar as possible with the help of an assigned tool. Note that we rewrote each prompt as the initial prompt for the participant to start with by removing all keywords related to the image's style. \responseline{This is to provide a similar starting point for participants with different levels of expertise. However, participants are not required to use the initial prompt to start creation.} Table~\ref{tab:close-ended-tasks} in the Appendix shows the details of each close-ended study task.

\subsubsection{Protocol}
\label{subsubsec:protocol}

Each user study session took about 1.5 hours. At the beginning of each session, we asked participants for their consent to record. Participants were then assigned three tasks about text-to-image creations, each of them to be completed with either \textbf{\tool}, \textbf{Baseline}, or \textbf{Promptist}. To mitigate the learning effect, both task assignment order and tool assignment order were counterbalanced across participants. Before starting each task, one of the authors conducted an online tutorial to walk through and explain the features of the assigned tool. Then, participants were given a 5-minute practice period to familiarize themselves with the tool, followed by a 15-minute period to use the assigned tool to iteratively refine their image creations. After completing each task, participants filled out a post-task survey to give feedback about what they liked or disliked. Participants were also asked to answer five NASA Task Load Index (TLX) questions~\cite{hart1988development} as a part of the post-task survey. After completing all three tasks, participants filled out a final survey, where they directly compared three assigned tools. At the end of the study session, each participant received a \$25 Amazon gift card as compensation for their time.

\subsection{Results}
\label{subsec:results_1}

In this section, we report and analyze the difference in participants' performance in Study 1 when using {\tool} and the two baseline tools. For brevity, we denote the participants in the user study as P1-P12. 

\subsubsection{User Performance}
\label{subsubsec:close-end-results}

We measure a participant's performance in a close-ended task by calculating the \textit{structural similarity index} (SSIM)~\cite{wang2004image} between the generated image and the target image. We also refer readers to Fig.~\ref{fig:examples_1} in the Appendix for images generated by our participants. Table~\ref{tab:close-end} depicts participants' performance in three different tasks. Overall, participants using {\tool} yield the best performance on all three tasks, achieving an average SSIM of $0.648$. By contrast, participants using Baseline and Promptist achieved average SSIM of $0.479$ and $0.574$, respectively. We further used Welch’s $t$-test to examine the performance difference between different tools in each task. For task 1, the performance between the participants using {\tool} and the participants using Baseline is statistically different ($p<0.01$). However, the performance between the participants using {\tool} and the participants using Promptist is not statistically different ($p=0.49$). Our further investigation shows that this task is relatively simpler (which only includes one subject in the image) compared with the other two tasks (which include multiple subjects). The automatically refined prompt could already lead to a generated image that looks close to the target image. Therefore, it is not surprising that participants using Promptist also performed well in this task. Nevertheless, we found participants using {\tool} performed significantly better in Task 2 and Task 3 (both tasks include multiple objects in the images) compared with participants using Baseline ($p<0.01$ and $p=0.03$) and Promptist ($p=0.04$ and $p=0.03$). {\responseref{}Participants also self-rated how similar their generated image looks compared with the target image on a 7-point Likert scale (1---Dissimilar, 7---Similar). We found that participants using {\tool} performed significantly better (median rating: 6) compared with Baseline (median rating: 3.5, Wilcoxon signed-rank test: $p=0.02$) and Promptist (median rating: 5, Wilcoxon signed-rank test: $p=0.05$).}

To understand why participants using {\tool} performed better in the two challenging tasks, we analyzed the post-task survey responses and the recordings. We found that {\tool} users' better performance could be attributed to its \textit{multi-modal prompting support} and \textit{rich feedback}. First, we found participants heavily relied on exploring different modifiers in {\tool}'s prompting view. The average number of modifiers explored per participant is 4.8. By exploring different art styles ahead, participants were able to gain insights from {\tool} before generating an image (which usually takes about 30 seconds per generation in our user study). P8 commented, ``\textit{Exploring different styles [in {\tool}] can help me understand them in the suggested prompts.}'' By contrast, P9 wrote, ``\textit{I can not choose any art style from the model-refined prompt since I do not have enough expert knowledge [when using Promptist].}'' Second, participants also found the attention adjustment helpful during the image refinement. Based on the recordings, the average number of attention adjustments per participant was 4.3. In the post-task survey, all participants marked the attention adjustment as helpful. P9 said, ``\textit{[{\tool}] can precisely catch the parts that I want to improve through attention adjustment.}'' Meanwhile, participants using baselines struggled with aligning the model's attention with their creative intention. P7 complained, ``\textit{sometimes, the generated image does not include the elements that I wrote in the prompt [when using Baseline].}'' 

{\responseref{} We also further analyzed the impact of the initial prompt in Study 1. We found that 4 out of 12 participants did not use the provided initial prompt. Overall, these four participants achieved an average SSIM of $0.647$, which is similar to the overall performance among 12 participants (average SSIM: $0.648$). Therefore, we believe the initial prompt may have little effect on user performance.}

\subsubsection{User Ratings of Individual Features}
\label{subsubsec:feature-ratings-1}

\begin{figure*}
    \centering
    \includegraphics[width=\linewidth]{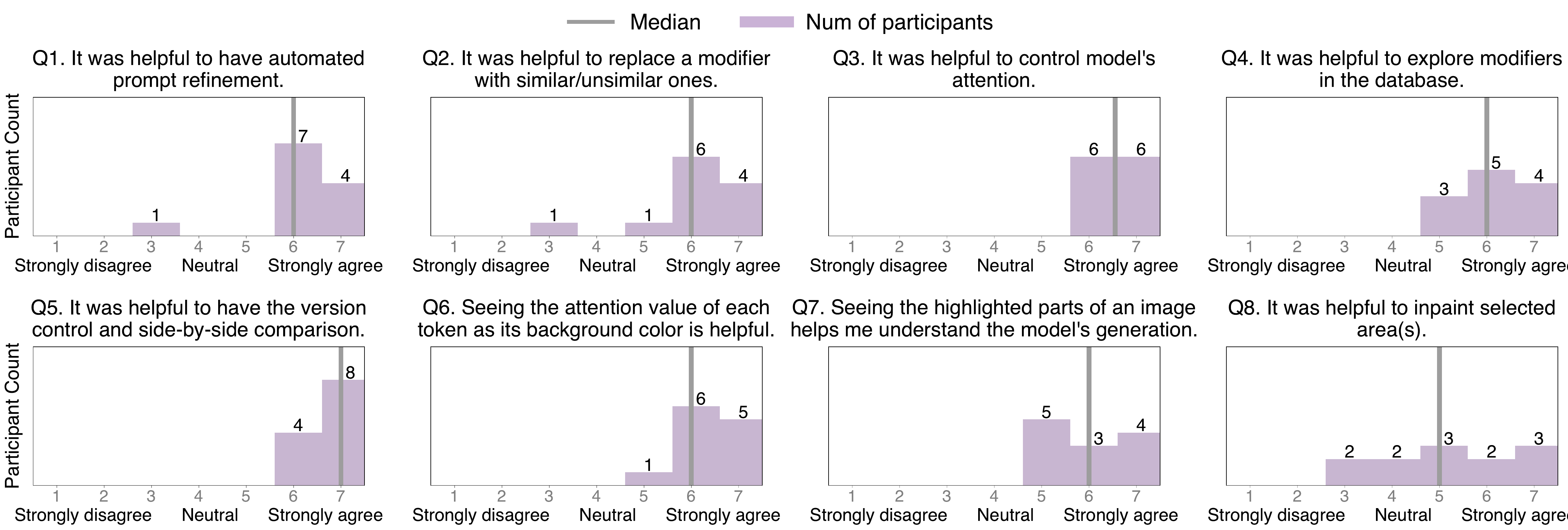}
    \caption{User's ratings on individual features of {\tool} in Study 1.}
    \Description{This figure shows the participants’ ratings about tool features of PromptCharm on a 7-point Likert scale (1 means strongly disagree and 7 means strongly agree) in Study 1. The median rating of automated prompt refinement is 6. The median rating of replacing modifiers is 6. The median rating of attention adjustment is 6.5. The median rating of image style exploration is 6. The median rating of version control is 7. The median rating of attention visualization is 6. The median rating of image highlighting is 6. The median rating of inpainting is 5.}
    \label{fig:feture_ratings_1}
\end{figure*}

In the post-task survey, 9 out of 12 participants indicated that they would like to use {\tool} for image creation in the future, while 3 participants stayed neutral. The median rating is 6 on a 7-point Likert scale (1---I do not want to use it at all, 7---I will definitely use it). Participants also rated the key features of {\tool} in the post-task survey. As shown in Fig.~\ref{fig:feture_ratings_1}, participants agreed most of the interactive features in {\tool} (median rating $\geq$ 6) were helpful except stayed neutral about \textit{image inpainting} (median rating: 5). The most appreciated features in {\tool} is the \textit{model attention adjustment}, where the median rating is 6.5. P12 commented, ``\textit{[{\tool}] can help me get better images as I imagined by adjusting the attention of different words in the prompt.}'' Besides, all participants also agreed that ``\textit{it was helpful to have the version control and side-by-side comparison.}'' 11 out of 12 participants agreed that ``\textit{seeing the attention value of each token as its background color was helpful.}''  The median rating is 6. P8 said, ``\textit{{\tool} can provide the information about the effect of each token [through attention visualization]. Then I know how to adjust the attention.}'' \responseline{Regarding \textit{image inpainting}, participants highlighted the need of performance improvement. For example, when trying to inpaint a relatively large area in the image, P11 found that the inpainted content did not fit the original image's background well. P11 thus commented, ``\textit{while generally it is acceptable, sometimes the inpainting effect is not ideal.}'' }

\subsubsection{Cognitive Overhead}
\label{subsubsec:cognitive-load-1}

\begin{figure}[t]
    \centering
    \includegraphics[width=0.95\linewidth]{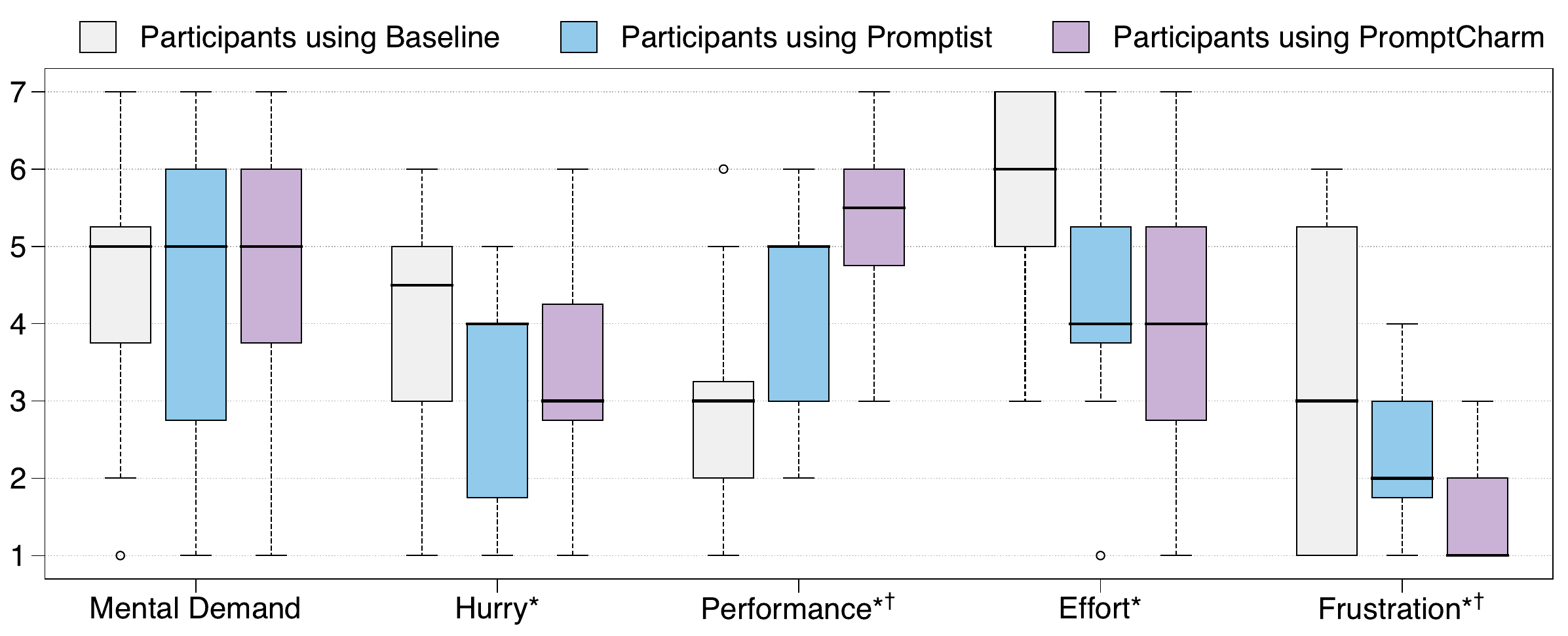}
    \caption{NASA Task Load Index Ratings in Study 1. Entries with a star (*) mean the mean difference is statistically significant ($p<0.05$) between {\tool} and Baseline.  Entries with a dagger ($\dag$) mean the mean difference is statistically significant ($p<0.05$) between {\tool} and Promptist.}
    \Description{This figure shows the participant’s ratings on 5 NASA task load indexes in Study 1. Participants using PromptCharm felt they had significantly better performance while experiencing significantly less frustration compared with participants using Baseline and Promptist. Besides, participants using PromptCharm also felt they spent significantly less effort compared with using Baseline. However, the differences in terms of mental demand and hurry among the three tools were not significant.}
    \label{fig:cognitive_load_1}
\end{figure}

{\responseref{}
Fig.~\ref{fig:cognitive_load_1} shows participants' ratings on the five cognitive factors of the NASA TLX questionnaire. Though {\tool} has more interactive features, we did not find statistical evidence indicating that participants using {\tool} experienced more mental demand compared with the participants using Baseline (Wilcoxon signed-rank test: $p=0.47$) and Promptist ($p=0.63$). However, participants using {\tool} felt they had better performance while experiencing less frustration compared with the participants using Baseline (Wilcoxon signed-rank test: $p<0.01$, $p<0.01$) and Promptist (Wilcoxon signed-rank test: $p=0.04$, $p=0.03$). In terms of effort and hurry, participants felt they experienced less hurry and spent less effort when using {\tool} compared with using Baseline (Wilcoxon signed-rank test: $p=0.03$ and $p<0.01$). However, such differences between using {\tool} and Promptist are not significant (Wilcoxon signed-rank test: $p=0.61$ and $p=0.22$). 
}

\iffalse
Fig.~\ref{fig:cognitive_load_1} shows participants' ratings on the five cognitive factors of the NASA TLX questionnaire. Though {\tool} has more interactive features, we did not find statistical evidence indicating that participants using {\tool} experienced more mental demand or hurry compared with the participants using Baseline (Welch's $t$-test: $p=0.5$, $p=0.13$) and Promptist ($p=0.54$, $p=0.70$). However, participants using {\tool} felt they had better performance while experiencing less frustration compared with the participants using Baseline (Welch's $t$-test: $p<0.01$, $p<0.01$) and Promptist (Welch's $t$-test: $p=0.04$, $p=0.04$). In terms of effort, participants felt they spent less effort when using {\tool} compared with using Baseline (Welch's $t$-test: $p=0.01$). However, such difference between using {\tool} and Promptist is not significant (Welch's $t$-test: $p=0.29$). 
\fi

\begin{figure*}[t]
    \centering
    \includegraphics[width=\linewidth]{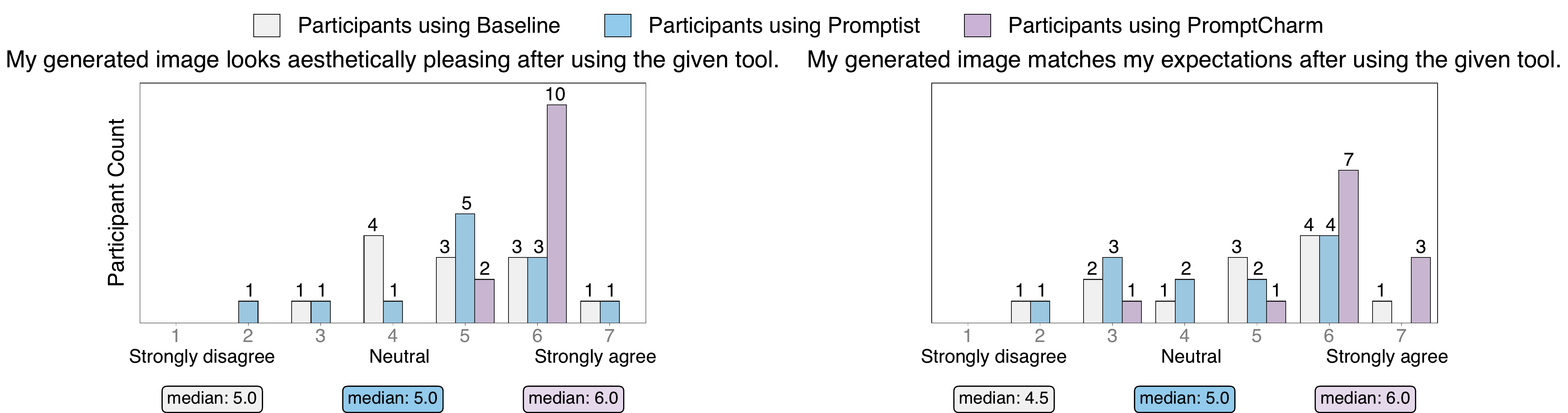}
    \caption{User's self-assessments of their performance in Study 2.}
    \Description{This figure shows the participants’ self-assessments regarding their performance in Study 2 from different perspectives: aesthetically pleasing and matching expectations. In terms of aesthetically pleasing, the median rating of Baseline, Promptist, and PromptCharm users are 5, 5, and 6. In terms of matching expectations, the median ratings of Baseline, Promptist, and PromptCharm users are 4.5, 5, and 6.}
    \label{fig:performance_2}
\end{figure*}

\subsubsection{User Preference and Feedback}
\label{subsubsec:preferences-feedback-1}

In the final survey, participants self-reported their preference among {\tool}, Baseline, and Promptist. All 12 participants found that {\tool} was the most helpful among three tools (median ranking: 1) and they preferred to use it in practice (median ranking: 1). We coded participants' responses in the final survey and found their preference mainly come from two sources. First, 9 out of 12 participants highlighted the contribution of the \textit{attention adjustment}. P11 wrote, ``\textit{in {\tool}, I can adjust the attention of each word and thereby refine my images directly.}'' Second, 8 out of 12 participants mentioned that \textit{image style exploration} helps them understand a particular style without actually generating an image. P10 commented, ``\textit{with {\tool}, I can easily know the reason why I choose an image style keyword even though I do not have such domain knowledge.}'' 

Participants also mentioned the limitations in the current version of {\tool} in the final survey. 3 out of 12 participants mentioned the performance of inpainting could be further improved. 2 out of 12 participants suggested adding an interactive feature that users can directly drag a subject to a specific position. 

\section{User Study 2: Open-ended Tasks}
\label{sec:user_study_2}

We conducted another within-subjects user study with another 12 participants to evaluate {\tool}'s usability when performing open-ended tasks. We compared {\tool} with the same baselines (Baseline and Promptist) as Study 1 (Sec.~\ref{sec:user_study_1}). We also followed the same protocol as Study 1 (Sec.~\ref{subsubsec:protocol}).

\subsection{Methods}
\label{subsec:methods_2}

\subsubsection{Participants}
\label{subsubsec:participants_2}

We recruited 12 participants through mailing lists of the ECE and CS departments at a research university~\footnote{This human-participated study is approved by the university’s research ethics office.}. 6 participants were Master students, 6 were Ph.D. students. Participants were asked to self-report their experience with general generative AI (e.g., ChatGPT) and text-to-image creation models (e.g., DALL-E and Midjourney). Regarding general generative AI, 1 participant had 2-5 years of experience, 7 had 1 year, and 4 had less than 1 year or no experience. Regarding text-to-image creation experience, 3 had 1 year of experience, and 9 had less than 1 year or no experience. All participants mentioned that they had never used any prompt engineering tools before. We conducted all user study sessions via Zoom. {\tool} and two baselines were again deployed as web applications. Therefore, participants were able to access our study sessions from their own PCs. 

\subsubsection{Tasks}
\label{subsubsec:tasks_2}
Different from Study 1, Study 2 does not assign the participants with the target images. For each task, we provided a participant with subject(s) to be included in the image. We designed three different image scenes and each of them features one or multiple subjects. The participants were asked to create an image in each task given one of the three image scenes with the help of an assigned tool based on their own creative ideas. Though giving specific image subjects might limit a participant's creative freedom, it yielded a more controlled experiment for assessing a tool's usability. Table~\ref{tab:open-ended-tasks} in the Appendix shows the details of each open-ended task in our user study.

\subsection{Results}
\label{subsec:results_2}

In this section, we report and analyze the difference in participants' performance in Study 2 when using {\tool} and the two baseline tools. For brevity, we denote the participants in Study 2 as P13-P24 to distinguish them from Study 1. 

\begin{figure*}[t]
    \centering
    \includegraphics[width=\linewidth]{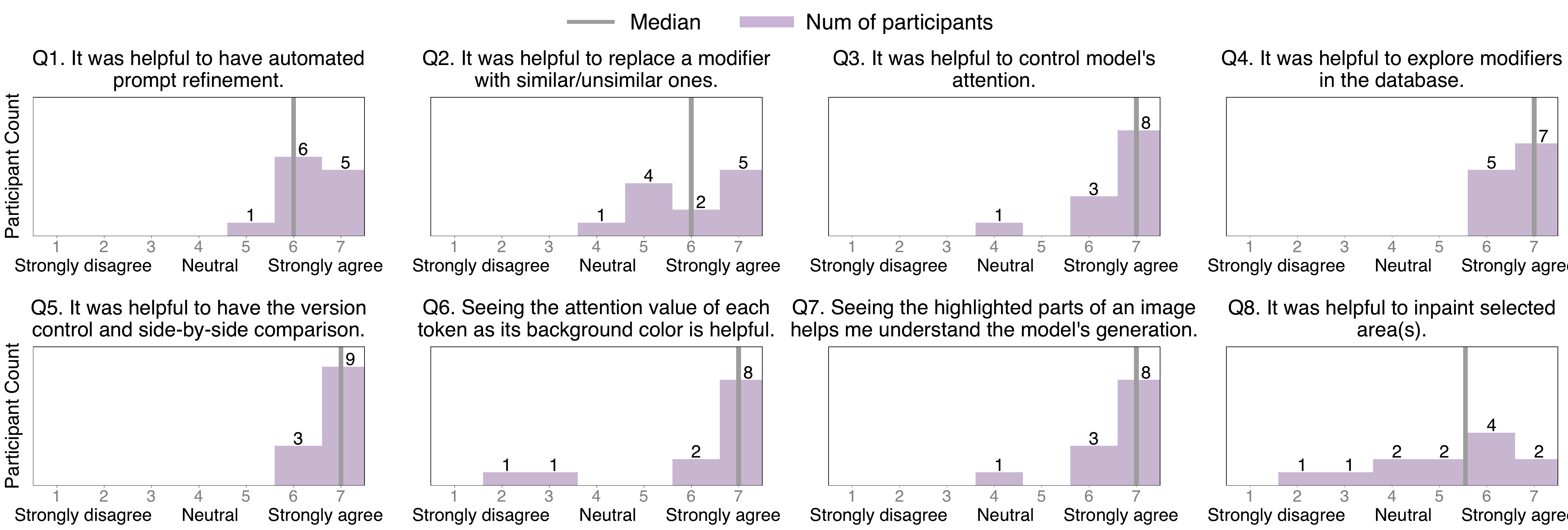}
    \caption{User's ratings on individual features of {\tool} in Study 2.}
    \Description{This figure shows the participants’ ratings about tool features of PromptCharm on a 7-point Likert scale (1 means strongly disagree and 7 means strongly agree) in Study 2. The median rating of automated prompt refinement is 6. The median rating of replacing modifiers is 6. The median rating of attention adjustment is 7. The median rating of image style exploration is 7. The median rating of version control is 7. The median rating of attention visualization is 7. The median rating of image highlighting is 7. The median rating of inpainting is 5.5.}
    \label{fig:feture_ratings_2}
\end{figure*}

\subsubsection{User Performance}
\label{subsubsec:open-end-results}

As there is no objective measurement for user performance in open-ended tasks, we asked participants to self-report their assessments to the quality of their generated images in the post-task survey in Study 2. We also refer readers to Fig.~\ref{fig:examples_2} in the Appendix for images generated by our participants. In the post-task survey, we asked participants to answer the following two 7-point Likert scale (1---Strongly disagree, 7---Strongly agree) questions: (1) \textit{my generated image looks aesthetically pleasing after using the assigned tool}, and (2) \textit{my generated image matches my expectations after using the assigned tool}. Fig.~\ref{fig:performance_2} shows participants' assessments when using {\tool} and two baseline tools. We found that participants felt their generated images were more aesthetically pleasing (median rating: 6) and closer to their expectations (median rating: 6) when using {\tool} compared with using either Baseline (median ratings: 5 and 4.5) or Promptist (median ratings: 5 and 5). In terms of aesthetically pleasing, there is a statistically significant performance difference between the participants using {\tool} and Baseline (mean difference: $5.8$ vs. $4.9$, \responseline{Wilcoxon signed-rank test: $p=0.02$}), as well as between {\tool} and Promptist (mean difference: $5.8$ vs. $4.9$, \responseline{Wilcoxon signed-rank test: $p<0.01$}). In terms of matching expectations, there is also a statistically significant performance difference between {\tool} and Baseline (mean difference: $5.9$ vs. $4.4$, \responseline{Wilcoxon signed-rank test: $p=0.02$}), as well as between {\tool} and Promptist (mean difference: $5.9$ vs. $4.8$, \responseline{Wilcoxon signed-rank test: $p=0.04$}). 

{\responseref{} 
We further analyzed participants' qualitative responses in the post-task survey and the video recordings and compared the results with Study 1. While similar features such as \textit{exploring modifiers} and \textit{model attention adjustment} had contributed to the success, we found participants in Study 2 shared different user behaviors. First, when exploring image modifiers, participants in Study 1 usually ended up selecting only two or three keywords. Then, they would try out different combinations and replacements to make the generated image look closer and closer to the target one. By contrast, participants in Study 2 explored and selected many more modifiers (7.3 modifiers on average) and experimented with more different image styles. As a result, 8 out of 12 participants explicitly mentioned in the post-task survey that {\tool} enables them to discover a diverse set of image styles. P22 commented, ``\textit{[{\tool}] gives me a lot of different image style suggestions. I can generate my image more freely.}'' By contrast, participants using two baselines had a hard time selecting image styles. P15 said, ``\textit{[when using Baseline,] it was easy for me to describe an image scene. But it was hard for me to provide keywords about image styles.}'' Second, since participants were given more freedom in Study 2, we found that they tended to write longer and more complex prompts. In this case, with the help of  {\tool}'s model attention adjustment, participants did not need to worry about the model losing attention to subjects after appending a lot of image modifiers. For instance, P20 wrote a long prompt with more than 60 tokens for Task 1 and found that the generated image missed the object ``cat.'' Such a problem was then easily fixed by P20 with model attention adjustment to the word ``cat.'' P20 then commented, ``\textit{[attention adjustment] makes things easier. I can now focus on selecting image styles.}''
}

% We further analyzed participants' qualitative responses in the post-task survey and the video recordings to interpret such performance differences. First, 8 out of 12 participants explicitly mentioned in the post-task survey that exploring different modifiers in {\tool} enables them to discover a diverse set of image styles. P13 commented, ``\textit{[with {\tool},] I can easily understand and select the image style that I want the model to generate.}'' By contrast, participants using two baselines had a hard time selecting image styles. P15 said, ``\textit{[when using Baseline,] it was easy for me to describe an image scene. But it was hard for me to provide keywords about image styles.}'' Second, participants heavily relied on two image refinement methods (i.e., model attention adjustment and image inpainting) when using {\tool}. The average numbers of attention adjustment and inpainting per participant are 4.6 and 2.1, respectively. P17 wrote, ``\textit{the way to adjust attention score of words can make the image looks more ideal.}'' However, participants using two baselines could experience frustration when refining their images. P18 complained, ``\textit{[when using Promptist,] I was more satisfied with the style of a new image. However, I could not inpaint the small undesired areas. In the end I had to choose another image which I disliked its style but did not introduce extra subjects.}''

\subsubsection{User Ratings of Individual Features}
\label{subsubsec:feature-ratings-2}

11 out of 12 participants indicated that they would like to use {\tool} for image creation in the future in the post-task survey, while 1 participant stayed neutral. The median rating is 6 on a 7-point Likert scale (1---I do not want to use it at all, 7---I will definitely use it). 
Regarding the individual features, all participants agreed that \textit{it was helpful to explore modifiers in the database} (median rating: 7). P16 wrote, ``\textit{[in {\tool}], I can easily understand the meaning of each keyword by exploring them.}'' All participants also agreed that the \textit{version control} was helpful (median rating: 7). Furthermore, 11 out of 12 participants also agreed that \textit{it was helpful to control model's attention}. P13 commented, ``\textit{I can simplify the complex prompting work by directly adjusting the attention.}''

\subsubsection{Cognitive Overhead}
\label{subsubsec:cognitive-load-2}

\begin{figure}[t]
    \centering
    \includegraphics[width=0.95\linewidth]{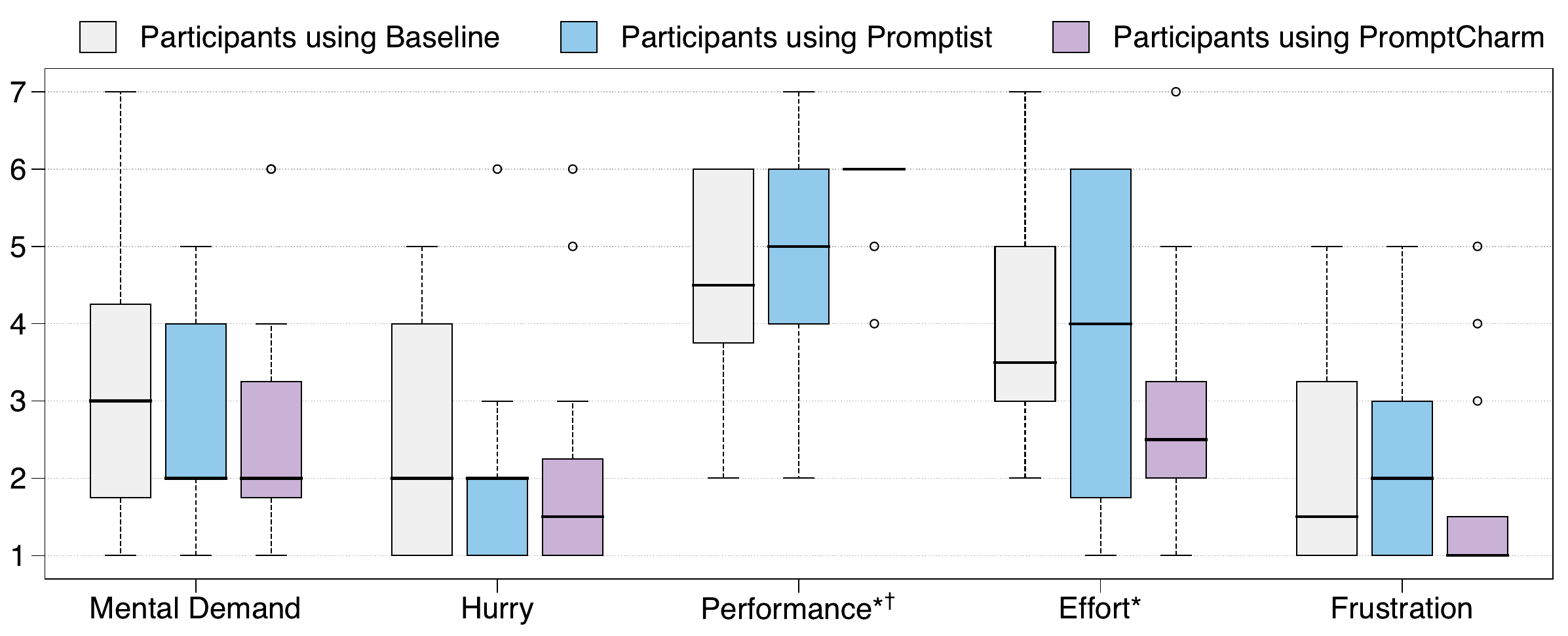}
    \caption{NASA Task Load Index Ratings in Study 2. Entries with a star (*) mean the mean difference is statistically significant ($p<0.05$) between {\tool} and Baseline.  Entries with a dagger ($\dag$) mean the mean difference is statistically significant ($p<0.05$) between {\tool} and Promptist.}
    \Description{This figure shows the participant’s ratings on 5 NASA task load indexes in Study 2. Participants using PromptCharm felt they had significantly better performance compared with participants using Baseline and Promptist. Besides, participants using PromptCharm also felt they spent significantly less effort compared with using Baseline. However, the differences in terms of mental demand, hurry, and frustration among the three tools were not significant.}
    \label{fig:cognitive_load_2}
\end{figure}

Fig.~\ref{fig:cognitive_load_2} presents participants' assessments to the five cognitive factors of the NASA TLX questionnaire in Study 2. As shown in Fig.~\ref{fig:cognitive_load_2}, we did not find significant differences between {\tool} and two baselines regarding mental demand (\responseline{Wilcoxon signed-rank test: $p=0.21$, $p=0.42$}), hurry (\responseline{Welch's $t$-test: $p=0.25$, $p=0.55$}), and frustration (\responseline{Wilcoxon signed-rank test: $p=0.18$, $p=0.19$}). These results indicate that more interaction in {\tool} did not introduce additional burdens or learning barriers to users in open-ended tasks. Moreover, when using {\tool}, participants significantly felt they had better performance compared with using Baseline (\responseline{Wilcoxon signed-rank test: $p=0.01$}) and Promptist (\responseline{Wilcoxon signed-rank test: $p=0.02$}). Participants also felt they were spending less effort when using {\tool} compared with using Baseline (\responseline{Wilcoxon signed-rank test: $p=0.04$}), while such difference is not significant between {\tool} and Promptist (\responseline{Wilcoxon signed-rank test: $p=0.18$}). \responseline{We also found that Study 2's participants experienced less mental demand and spent less effort compared with Study 1's participants (Fig.~\ref{fig:cognitive_load_1}). A plausible explanation is participants might feel less stressed when performing the open-ended tasks (Study 2).}

\subsubsection{User Preference and Feedback}
\label{subsubsec:preferences-feedback-2}

In the final survey, 11 out of 12 participants found that {\tool} was the most helpful among three conditions (median ranking: 1). Besides, all participants indicated that they preferred to use {\tool} more compared with the two baselines (median raking: 1). After coding the participants' responses, we identified two different themes that led to such preference. First, participants felt they were directly interacting with the model when using {\tool}. P18 said, ``\textit{I could modify the prompt in a way where it tells me I'm interacting directly with the model. While for the other two, I had to guess a lot of things without knowing where the model is putting more importance.}'' P16 commented, ``\textit{I can gain more control over the image generation process [when using {\tool}]. Thus, I felt more confident about how to generate an expected image.}'' Second, participants also appreciated the rich feedback loop in {\tool} as it yielded a higher efficiency. P21 wrote, ``\textit{[in {\tool}], I have more choices when I need to refine my image, e.g., I can either change the attention or directly erase the elements that I do not want. This is the most time-saving among the three tools.}'' P14 said, ``\textit{exploring different modifiers save my time. I can check different image styles without actually using the model to generate one.}''

Participants also gave their feedback about the potential improvements of {\tool} in the post-task and final surveys. In addition to those discussed in Study 1 (Sec.~\ref{subsubsec:preferences-feedback-1}), 2 out of 12 participants further mentioned the need for textual explanations for different image modifiers in {\tool}'s \textit{image style exploration}.

\section{Discussion}
\label{sec:discussions}

The rise of generative models and prompt engineering have significantly influenced many fields and domains. Our user study results indicate that even novice users can use the stable diffusion model to create aesthetically pleasing images with rich interactive prompt engineering support. Therefore, it is worth continuing to investigate new interaction designs to assist novice users in prompt engineering. This section discusses the implications derived from our system design and user studies, as well as limitations and future work.

% human-ai collaboration guidelines, mixed-initiative, 
\subsection{Assisting Effective Prompt Engineering with Model Explanations}
% \smalltitle{Prompting less with more feedback.} 
When designing human-AI interaction, an important guideline is \textit{making clear why the system did what it did}~\cite{amershi2019guidelines}. However, to the best of our knowledge, there is little investigation into how to provide model explanations for prompt engineering. During our user studies, we particularly found that the lack of proper model explanations may lead to user frustration. For example, when working on Task 3 in Study 2 without model explanations (Baseline), P18 commented, ``\textit{I could not figure out why the model keeps generating multiple penguins until I manually changed the word `playing' to `rolling'. Then I realized that the model might have a high attention on this word. I could have found this easily if I have any support like attention visualization.}'' %{\tool} provides the exact support of model attention visualization. 
When using {\tool}, by observing the model's attention to different keywords in the prompt, the user can quickly identify if the model has missed any keywords during the generation. As a result, the user can refine their prompts and images in a more targeted manner. For instance, they can directly increase the model's attention to a keyword if they found this keyword has been misattended by the model during the generation.
% Moreover, this guideline should likewise be followed when using other AI models. For instance, when talking about the automated prompt refinement in Promptist, P15 said, ``\textit{I have no ideas about the meaning of the words it added, so I do not know how to change the prompt to generate a picture that I want.}'' To address this, {\tool} supports \textit{image style exploration}. P15 continued to comment when using {\tool}, ``\textit{I found {\tool} addressed my need by supporting me explore the generated prompts with example pictures from the database.}'' 

\subsection{Enriching User Feedback Loop in Prompt Engineering}
Prompting has proven to be a valuable asset in human-AI collaboration, as large language models can now effectively comprehend natural language instructions from users and directly translate them into actions. However, when addressing specific tasks, such as text-to-image creation, relying solely on prompting as the interface between the user and generative models is insufficient. On the one hand, the model should help users understand how their feedback has been incorporated during the generation. On the other hand, the model should provide flexible ways to elicit user feedback. Ultimately, the goal is to create images that better align with the user's creative intent, rather than writing complex and long prompts. The design of {\tool} is highly influenced by one of the design principles for mixed-initiative user interfaces~\cite{horvitz1999principles}---\textit{providing mechanisms for efficient agent-user collaboration to refine results.} {\tool} enables multi-modal prompt refinement, including image style exploration, model attention visualization, attention adjustment, and image inpainting. According to our user study results, such mixed-initiative design led to better user performance in both close-ended and open-ended tasks.

\subsection{Addressing Conceptual Gaps}
Our user studies reveal significant \textit{conceptual gaps} when novice users write prompts for image creation. In many cases, users have a clear image style in mind, but they find it challenging to articulate it in a prompt. This difficulty becomes more pronounced when it comes to selecting effective keywords (i.e., modifiers) for the stable diffusion model. Broadly speaking, this issue resembles one of the six learning barriers in designing end-user programming systems---\textit{selection barriers}~\cite{ko2004six}, where \textit{users know what they want the computer to do, but they do not know what to use}. To address this, {\tool} supports \textit{image style exploration}. The user can efficiently browse popular image style keywords and further contextualize them by retrieving relevant images from a large database. As a result, participants performed better when using {\tool} to solve the close-ended tasks, which require them to replicate a particular image and its style. Such a design also facilitates users understanding the model-refined prompts. For instance, when talking about the automated prompt refinement, P15 said, ``\textit{initially, I have no ideas about the meaning of the words it added, so I won't know how to change the prompt to generate a picture that I want.}'' P15 continued to comment, ``\textit{however, I found {\tool} addressed my need by supporting me explore the generated prompts' keywords with example pictures from the database.}'' 

\subsection{Exploration vs.~Exploitation}
The open-ended tasks and the close-ended tasks in our user study can represent two modes of interaction: \textit{exploration} and \textit{exploitation}. In the exploitation mode,  users have a clear understanding of their objectives. During Study 1 (close-ended tasks), we observed that participants spent a considerable amount of time comparing their generated images with the target images before making adjustments to the prompts or refining the images. For instance, when they found the generated image's art style was obviously different from the target image, they first analyzed which keyword in the prompts had contributed to the distinct style by examining the model's attention. Once they identified such keyword(s), they proceeded to replace it with another keyword that better matched the target style by selecting modifiers in {\tool}. In the exploration mode, users usually do not have a clear objective. In Study 2 (open-ended tasks), participants often started by formulating a basic prompt that included the image scene they had in mind. Subsequently, they focused on the ideation of the image styles by exploring and combining different and diverse image styles in {\tool}. Once they determined their desired image style, the process of refining images transitioned back to the exploitation mode. Such observation of the different modes' user behavior is similar to the previous studies on how programmers interact with AI programming assistants in \textit{acceleration} mode and \textit{exploration} mode~\cite{barke2023grounded, mcnutt2023design}. Our study results further imply that {\tool} can support both modes effectively. As Louie et al.~\cite{louie2020novice} pointed out, the human-AI interface for creative design should empower the user whether they have a clear creative goal in mind or not. Therefore, in future work, it is important to take \textit{exploration-exploitation balancing} into account when designing interactive systems for prompt engineering and creative design.

% \smalltitle{Balancing automation and user control.} The merits of fully automating user needs versus the importance of preserving user control has been a long-standing debate in our research community~\cite{shneiderman1997direct}. Yet, {\tool} opts for the mixed-initiative design, aiming to strike a balance between the AI model and the user. In {\tool}, the user's control is preserved throughout various prompting and refinement stages. For instance, the user can always edit the model-refined prompts to infuse their own creative intent by exploring and selecting new image styles. Though the image is ultimately generated by the model, the user can control the generation through adjusting the model attention. When the model inpaints an image, the user can guide this process by providing text instructions. This also aligns with the previous studies' results, as the user highly value the balance between AI automation and user control during a human-AI co-creation process~\cite{yan2022flatmagic}.

{\responseref{}
\subsection{User Experience vs. User Performance}

{\tool} is designed for users with different levels of expertise, especially for those who had limited experience in text-to-image creation. During our user study, only 6 out of 24 participants had 1 year of experience with text-to-image creation models, while the rest of them had less than 1 year of experience. In Study 1, we found that 3 participants who had 1 year of experience had similar performance (average SSIM: 0.642) compared with the others (average SSIM: 0.649). In Study 2, 3 participants with more experience had the same performance in terms of aesthetically pleasing (median rating: 6) and expectation matching (median rating: 6) compared with the others. These results indicate that the effectiveness of {\tool} is not strongly correlated to user expertise.}

\subsection{Limitations and Future Work}
\label{subsec:limitations}

There are several limitations in our user study design and system in addition to those pointed out by our user study participants as described in Sec.~\ref{subsubsec:preferences-feedback-1} and Sec.~\ref{subsubsec:preferences-feedback-2}.

\smalltitle{User Study Design.} In our current open-ended study, we manually designed three image scenes for participants. However, though such a design provides a more controllable set-up for evaluation, this may limit the participants' creative freedom. In future work, the open-ended study can be improved with the free-form usage of {\tool}~\cite{feng2023promptmagician}. Then, to assess a user's performance, one can consider the expert evaluation. \responseline{Besides, {\tool} has only been evaluated with novice users. In future work, we plan to evaluate {\tool}'s effectiveness with experts in text-to-image creation.}

\smalltitle{Generalization Issue.} Though we have only evaluated {\tool} on the \textit{Stable Diffusion}, we believe the design of {\tool} can generalize to other open-source text-to-image generation models, e.g., CogView~\cite{ding2021cogview}, VQGAN+CLIP~\footnote{\href{https://github.com/nerdyrodent/VQGAN-CLIP}{https://github.com/nerdyrodent/VQGAN-CLIP}}. Evaluating {\tool}'s usability with different text-to-image models may worth investigating in future work. To reuse {\tool} for close-source models such as DALL-E~\footnote{\href{https://openai.com/dall-e-2}{https://openai.com/dall-e-2}} and Midjourney~\footnote{\href{https://www.midjourney.com}{https://www.midjourney.com}}, alternative design for \textit{attention adjustment} and \textit{model explanations} need to be proposed. For instance, an alternative approach of \textit{attention adjustment} could be adding corresponding instructions in the prompts. For \textit{model explanations}, one may consider model-agnostic XAI methods, e.g., LIME~\cite{ribeiro2016should} and SHAP~\cite{lundberg2017shap}.

\smalltitle{Large-scale Exploration.} In the current design, {\tool} can only assist users in examining one generated image per version. This is because the main goal of {\tool} is to help novice users iteratively improve their creation based on the previous generation. Nevertheless, in future work, {\tool} could be improved to support experienced users exploring a large batch of generation (e.g., 30 images per prompt) simultaneously. To achieve this, specific design for image layout, e.g., organizing images by clustering based on the colors~\cite{feng2023promptmagician} or semantic similarity~\cite{brade2023promptify} may be leveraged.

\smalltitle{Alternative Algorithms and Design.} In the current version of {\tool}, we use \textit{Promptist}~\cite{hao2022optimizing}, a GPT-2 based model fine-tuned through reinforcement learning for automated prompt refinement given its popularity and availability. However, other different prompt refinement methods, e.g., gradient-based prompt search~\cite{wen2023hard}, few-shot prompting with powerful LLMs (GPT4)~\cite{brade2023promptify} might yield better prompting results. \responseline{Besides, our current inpainting method sometimes did not provide the user intended results, especially when the inpainted area is large (Sec.~\ref{subsubsec:feature-ratings-1}). This may be improved with the alternative design and models, e.g., \textit{prompt-to-prompt}~\cite{hertz2022prompt}, a method that uses the cross-attention map to enable content-preserving inpainting with the stable diffusion model.}

\section{Conclusion}
\label{sec:conclusion}

In this paper, we present {\tool}, a mixed-initiative system that assists novice users in text-to-image generation through multi-modal prompting and refinement. {\tool} provides automated prompt refinement to help users optimize their input text prompt with the help of a SOTA model, \textit{Promptist}. The user can further improve their prompt by exploring different image styles and keywords within {\tool}. To support users in effectively refining their images, {\tool} provides model explanations through model attention visualization. Once the user notices any unsatisfactory parts in the image, they can refine it by adjusting the model's attention to keywords, or masking those areas and re-generate it through an image inpainting model. Lastly, {\tool} integrates the version control to enable users to easily track their creations. Through a user study of 12 participants including close-ended tasks, we found that participants using {\tool} can create images that closely resembled the given target images when compared to two baseline methods. In another user study of 12 participants featuring open-ended tasks, participants using {\tool} felt that their generated images were more aesthetically pleasing and better matched their expectations compared with using two baselines. In the end, we discuss the design implications from {\tool} and propose several interesting future research opportunities.

\begin{acks}
We would like to thank all anonymous participants in the user studies and anonymous reviewers for their valuable feedback. This work was supported in part by Amii RAP program, Canada CIFAR AI Chairs Program, the Natural Sciences and Engineering Research Council of Canada (NSERC No.RGPIN-2021-02549, No.RGPAS-2021-00034, and No.DGECR-2021-00019), as well as JST-Mirai Program Grant No.JPMJMI20B8, JSPS KAKENHI Grant No.JP21H04877, No.JP23H03372.
This work was also supported in part by Amazon Research Award and the National Science Foundation (NSF Grant ITE-2333736).
\end{acks}

\balance

% \section{Acknowledgments}

% Identification of funding sources and other support, and thanks to
% individuals and groups that assisted in the research and the
% preparation of the work should be included in an acknowledgment
% section, which is placed just before the reference section in your
% document.

% \section{Appendices}

% If your work needs an appendix, add it before the
% ``\verb|\end{document}|'' command at the conclusion of your source
% document.

% Start the appendix with the ``\verb|appendix|'' command:
% \begin{verbatim}
%   \appendix
% \end{verbatim}
% and note that in the appendix, sections are lettered, not
% numbered. This document has two appendices, demonstrating the section
% and subsection identification method.

\bibliographystyle{ACM-Reference-Format}
\bibliography{reference}

\newpage
\appendix

\setcounter{figure}{0}
\def\thefigure{\Alph{section}\arabic{figure}}

\setcounter{table}{0}
\def\thetable{\Alph{section}\arabic{table}}

\def\thealgocf{\Alph{section}\arabic{algocf}}

\section{Study 1: Close-ended Tasks}

\subsection{Task Description}

\begin{table}[h]
    \centering
    \footnotesize
    \caption{Close-ended User Study Tasks}
    \begin{tabular}{l|l|l|l}
        \toprule
        \# & \multicolumn{1}{c|}{Task 1} & \multicolumn{1}{c|}{Task 2} & \multicolumn{1}{c}{Task 3} \\
        \midrule
         & \multicolumn{1}{c|}{\multirow{1}{*}{{\includegraphics[width=0.2\linewidth]{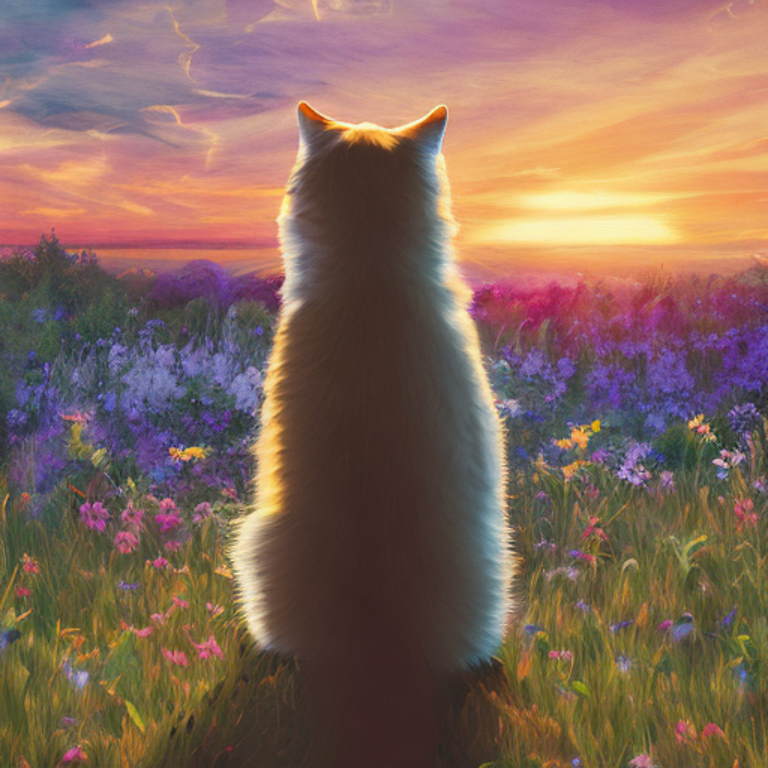}}}} & \multicolumn{1}{c|}{\multirow{1}{*}{{\includegraphics[width=0.2\linewidth]{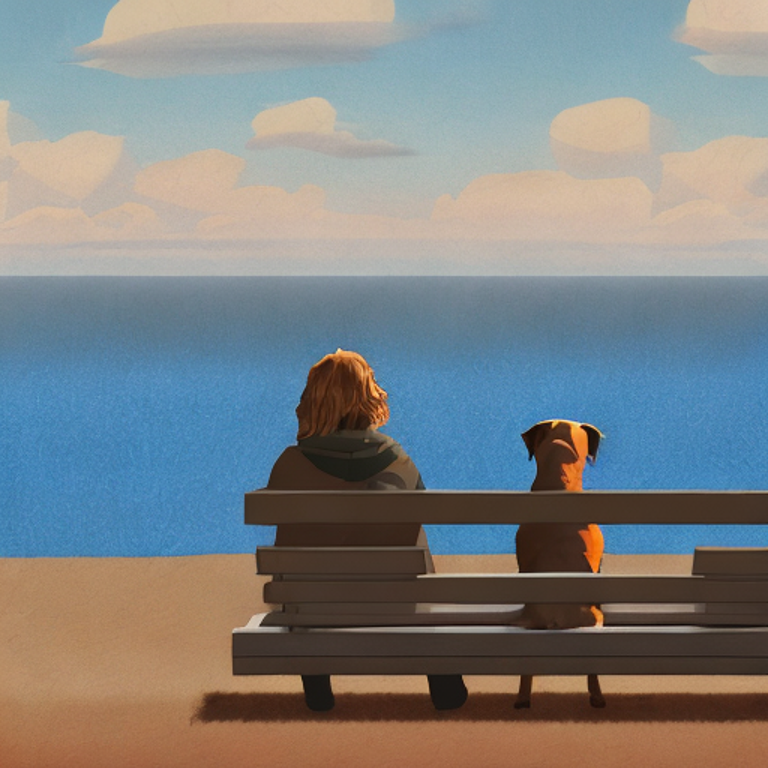}}}} & \multicolumn{1}{c}{\multirow{1}{*}{{\includegraphics[width=0.2\linewidth]{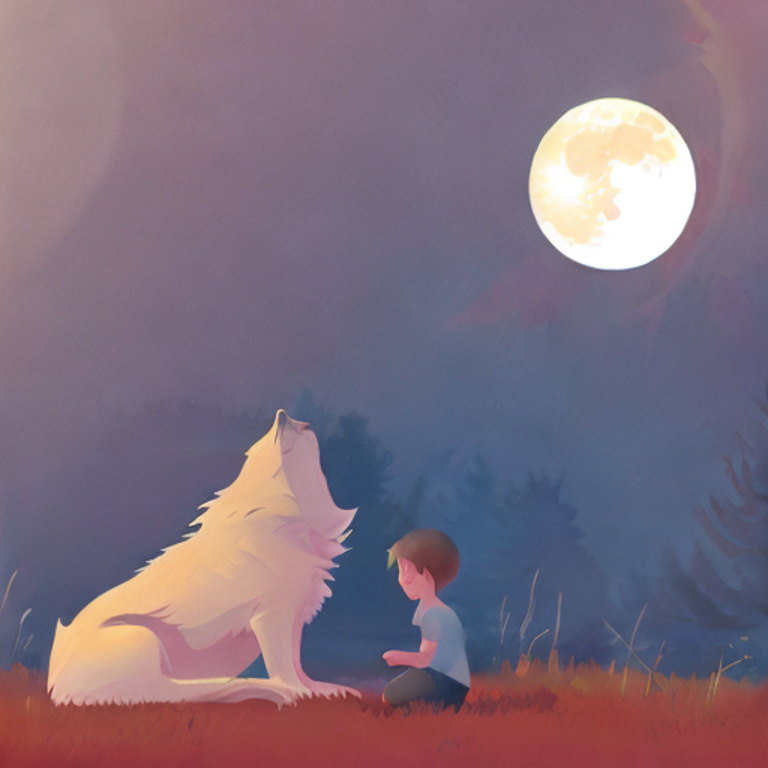}}}} \\
        & & & \\
        Target & & & \\
        Image & & & \\
        & & & \\
        & & & \\
        \midrule
        & A painting from & An image of a girl & A painting of a wolf \\
        & behind of an orange & and her dog sitting & sitting next to a \\
        Initial & cat standing in a & on a bench and & human child in front \\
        Prompt & field of flowers and & looking at the sea. & of the full moon.\\
        & watching the & & \\
        & sunset. & & \\
        % 1 & A painting from behind of an orange cat standing in& \multirow{1}{*}{{\includegraphics[width=0.2\textwidth]{close_study/gt1.png}}} \\
        %   & a field of flowers and watching the sunset.& \\
        %   & & \\
        %   & & \\
        %   & & \\
        %   & & \\
        %   & & \\
        %   & & \\
        % \midrule
        % 2 & An image of a girl and her dog sitting on a bench and& \multirow{1}{*}{{\includegraphics[width=0.2\textwidth]{close_study/gt2.png}}} \\
        %   & looking at the sea.& \\
        %   & & \\
        %   & & \\
        %   & & \\
        %   & & \\
        %   & & \\
        %   & & \\
        % \midrule
        % 3 & A painting of a wolf sitting next to a human child in& \multirow{1}{*}{{\includegraphics[width=0.2\textwidth]{close_study/gt3.png}}} \\
        %   & front of the full moon.& \\
        %   & & \\
        %   & & \\
        %   & & \\
        %   & & \\
        %   & & \\
        %   & & \\
        \bottomrule
    \end{tabular}
    \label{tab:close-ended-tasks}
\end{table}

\subsection{Examples of User-created Images}

\begin{figure}[h]
    \centering
    \includegraphics[width=0.95\linewidth]{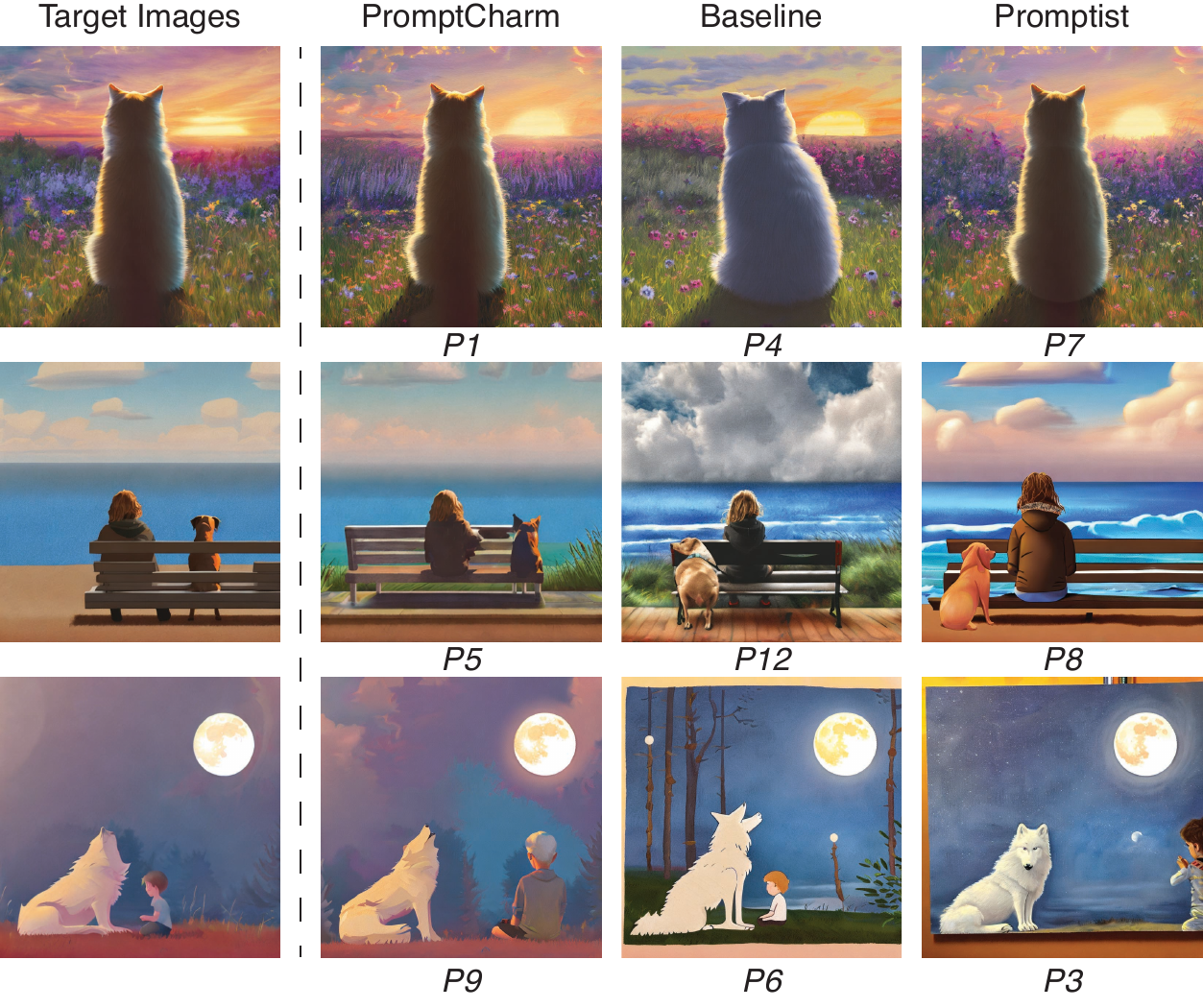}
    \caption{Examples of user-created images in Study 1}
    \Description{This figure includes Study 1 participants’ creations. The left column includes three target images. And the right three columns include PrompCharm, Baseline, and Promptist results, respectively.}
    \label{fig:examples_1}
\end{figure}

\section{Study 2: Open-ended Tasks}

\subsection{Task Description}

\begin{table}[h]
    \footnotesize
    \centering
    \caption{Open-ended User Study Tasks}
    \begin{tabular}{l|l|l}
         \toprule
         \# & \multicolumn{1}{c|}{Image Subject(s)} & \multicolumn{1}{c}{Image Scene}  \\
         \midrule
         Task 1 & A dog and a cat & The dog is playing with the cat \\
         Task 2 & A group of Canada geese & The geese are walking on a street \\
         Task 3 & A penguin & The penguin is playing in the snow \\
         \bottomrule
    \end{tabular}
    \label{tab:open-ended-tasks}
\end{table}

\subsection{Examples of User-created Images}

\begin{figure}[H]
    \centering
    \includegraphics[width=\linewidth]{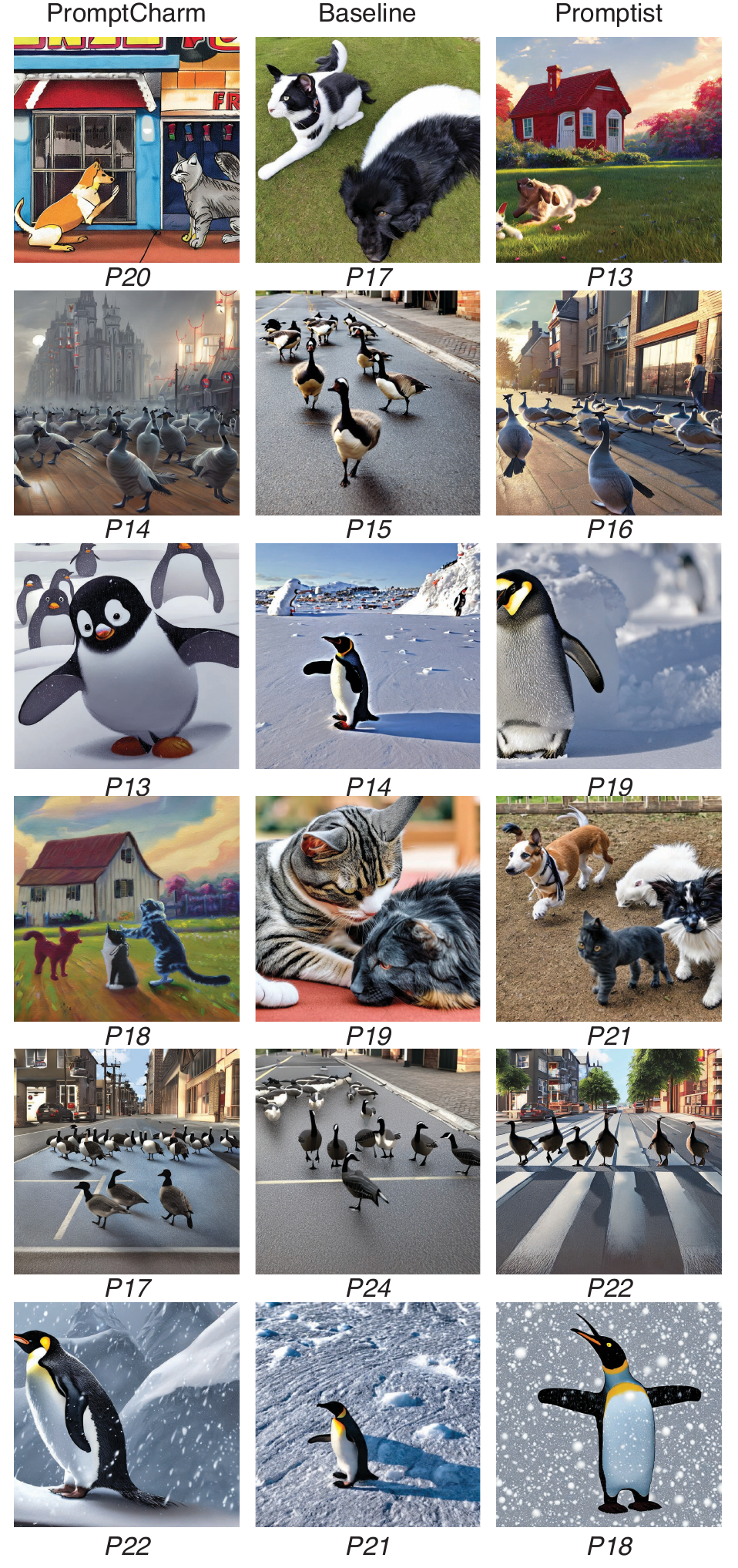}
    \caption{Examples of user-created images in Study 2}
    \Description{This figure includes Study 2 participants’ creations. The top part includes images in a 3x3 grid, where each column represents the examples of three different tasks created by PromptCharm, Baseline and Promptist. The bottom part of this figure presents another group of results.}
    \label{fig:examples_2}
\end{figure}

\newpage
{\responseref{}
\section{Model Attention Adjustment}

{\noindent \bf Example 1 prompt:}

{\ttfamily a clear image of a bustling city street at night, featuring a [taxi], a bus, neon signs, and people walking.} 

\begin{figure}[H]
     \centering
     \begin{subfigure}[t]{0.3\linewidth}
         \centering
         \includegraphics[width=\textwidth]{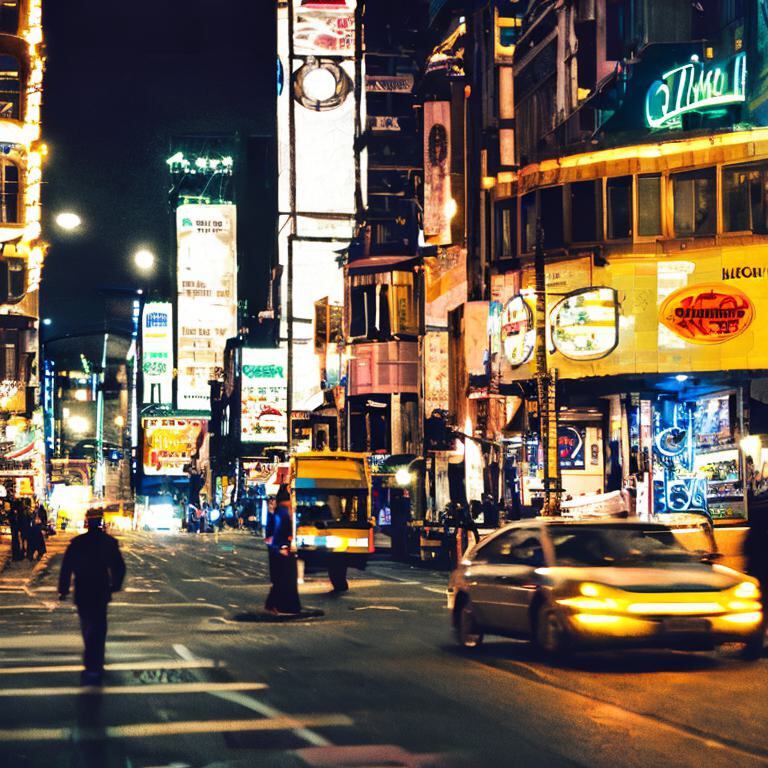}
         \caption{\responseref{}0.5x attention score to {\ttfamily [taxi]}}
         \label{fig:attention_a_1}
     \end{subfigure}
     \hfill
     \begin{subfigure}[t]{0.3\linewidth}
         \centering
         \includegraphics[width=\textwidth]{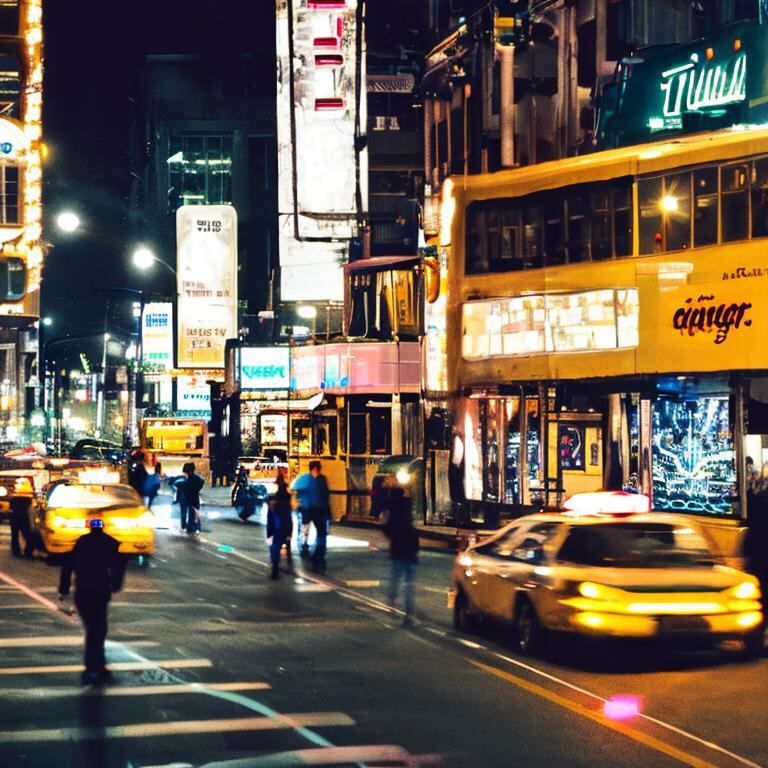}
         \caption{\responseref{}No adjustment.}
         \label{fig:attention_a_2}
     \end{subfigure}
     \hfill
     \begin{subfigure}[t]{0.3\linewidth}
         \centering
         \includegraphics[width=\textwidth]{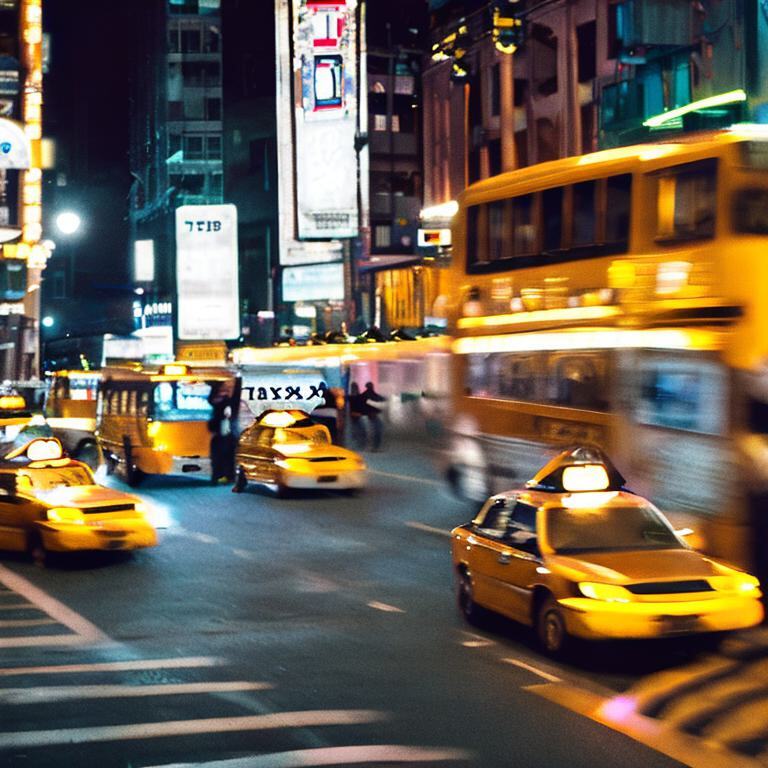}
         \caption{\responseref{}2x attention score to {\ttfamily [taxi]}}
         \label{fig:attention_a_3}
     \end{subfigure}
    \caption{\responseref{}Adjusting model attention to the word {\ttfamily [taxi]}.}
    \Description{This figure includes three different images. Image (a) has adjusted the model’s attention to [taxi] to 0.5, and the generated image does not include any taxis. Image (b) does not include any model attention adjustment, and the generated image includes one taxi. Image (c) has adjusted the model’s attention to [taxi] to 2, and the generated image includes multiple taxis.}
    \label{fig:attention_exp_1}
\end{figure}

{\noindent \bf Example 2 prompt:}

{\ttfamily a clear image of a cozy library room with a fireplace, comfortable armchairs, shelves filled with books, and a large window showing a [snowy] landscape outside.} 

\begin{figure}[H]
     \centering
     \begin{subfigure}[t]{0.3\linewidth}
         \centering
         \includegraphics[width=\textwidth]{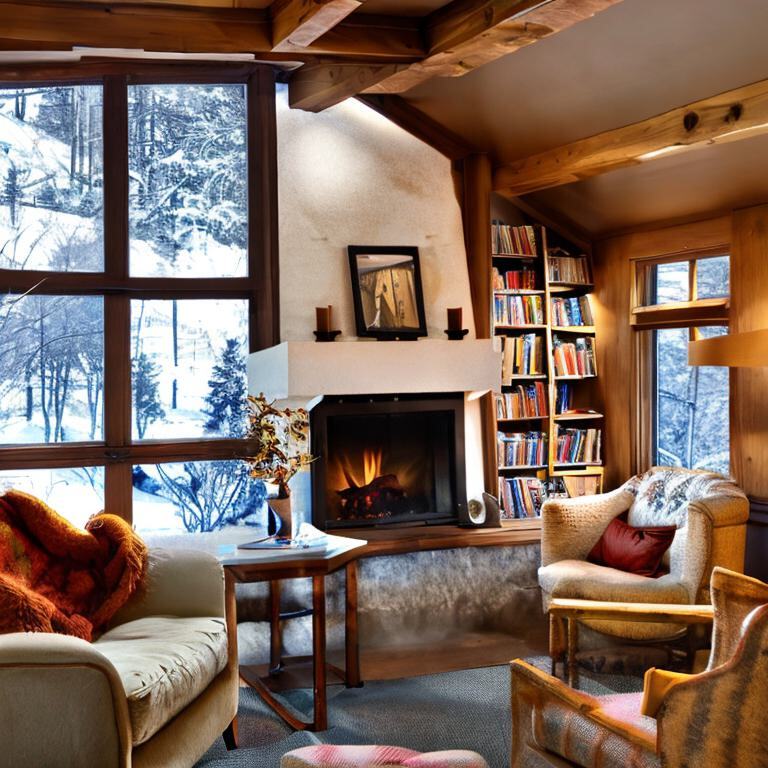}
         \caption{\responseref{}0.5x attention score to {\ttfamily [snowy]}}
         \label{fig:attention_b_1}
     \end{subfigure}
     \hfill
     \begin{subfigure}[t]{0.3\linewidth}
         \centering
         \includegraphics[width=\textwidth]{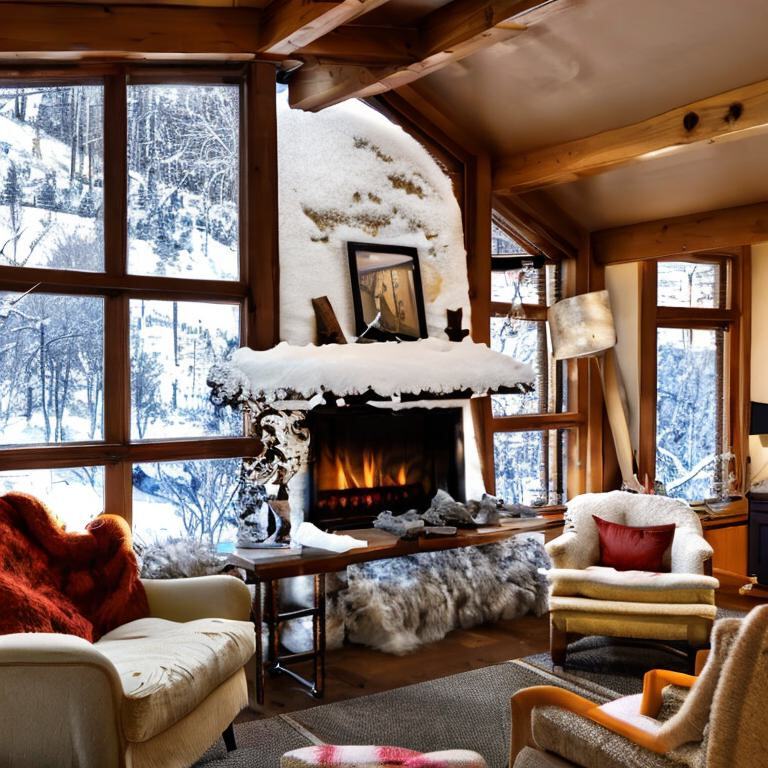}
         \caption{\responseref{}No adjustment.}
         \label{fig:attention_b_2}
     \end{subfigure}
     \hfill
     \begin{subfigure}[t]{0.3\linewidth}
         \centering
         \includegraphics[width=\textwidth]{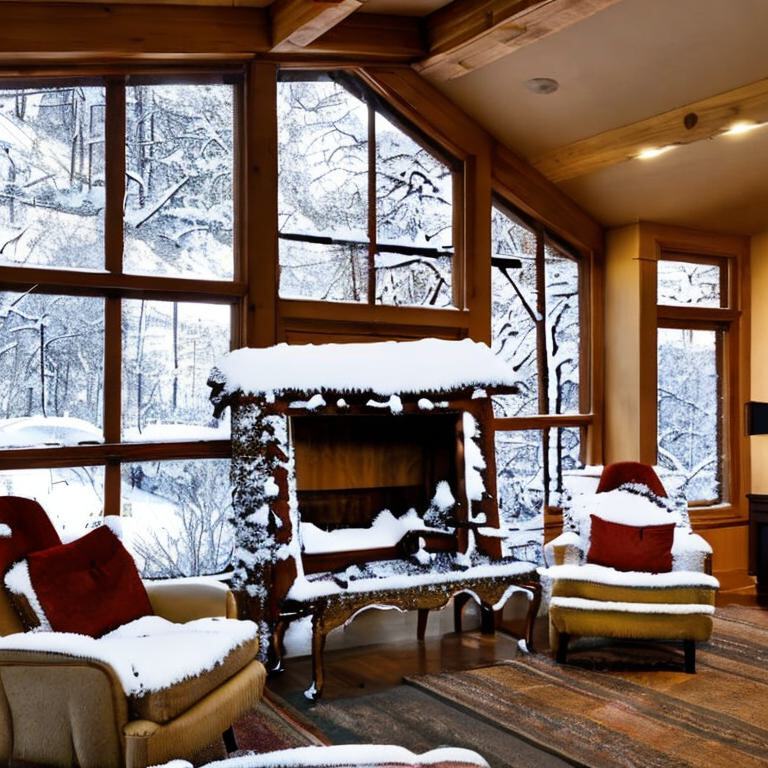}
         \caption{\responseref{}2x attention score to {\ttfamily [snowy]}}
         \label{fig:attention_b_3}
     \end{subfigure}
    \caption{\responseref{}Adjusting model attention to the word {\ttfamily [snowy]}.}
    \Description{This figure includes three different images. Image (a) has adjusted the model’s attention to [snowy] to 0.5, and the generated image does not include any snow. Image (b) does not include any model attention adjustment, and the generated image includes the fireplace covered by snow. Image (c) has adjusted the model’s attention to [snowy] to 2, and the generated image includes both sofas and the fireplace covered by snow.}
    \label{fig:attention_exp_2}
\end{figure}

% \newpage
{\noindent \bf Example 3 prompt:}

{\ttfamily an oil painting of a desert scene with cacti, under a heavy [rain], where each raindrop is distinctly visible.} 

\begin{figure}[H]
     \centering
     \begin{subfigure}[t]{0.3\linewidth}
         \centering
         \includegraphics[width=\textwidth]{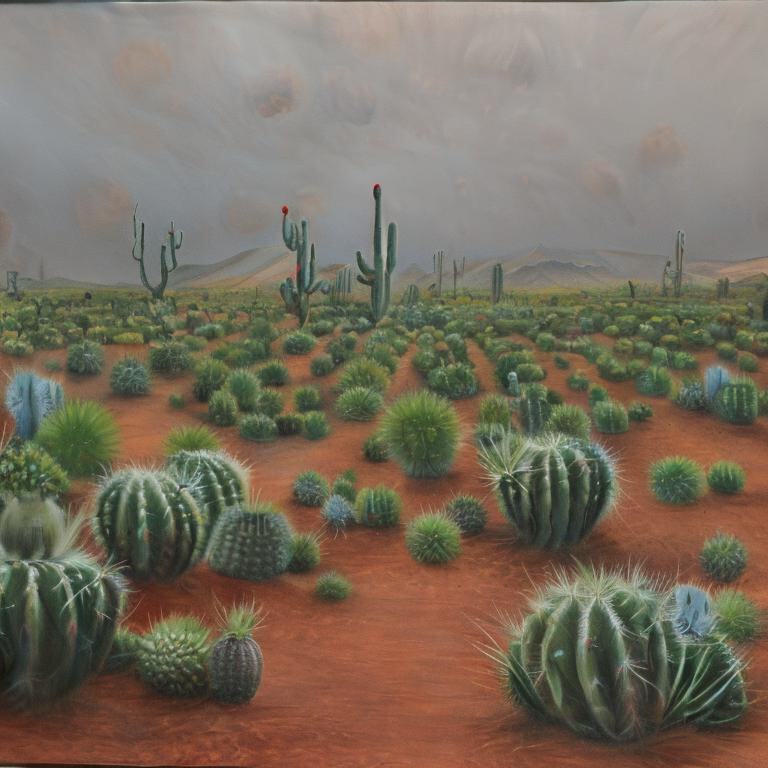}
         \caption{\responseref{}0.5x attention score to {\ttfamily [rain]}}
         \label{fig:attention_c_1}
     \end{subfigure}
     \hfill
     \begin{subfigure}[t]{0.3\linewidth}
         \centering
         \includegraphics[width=\textwidth]{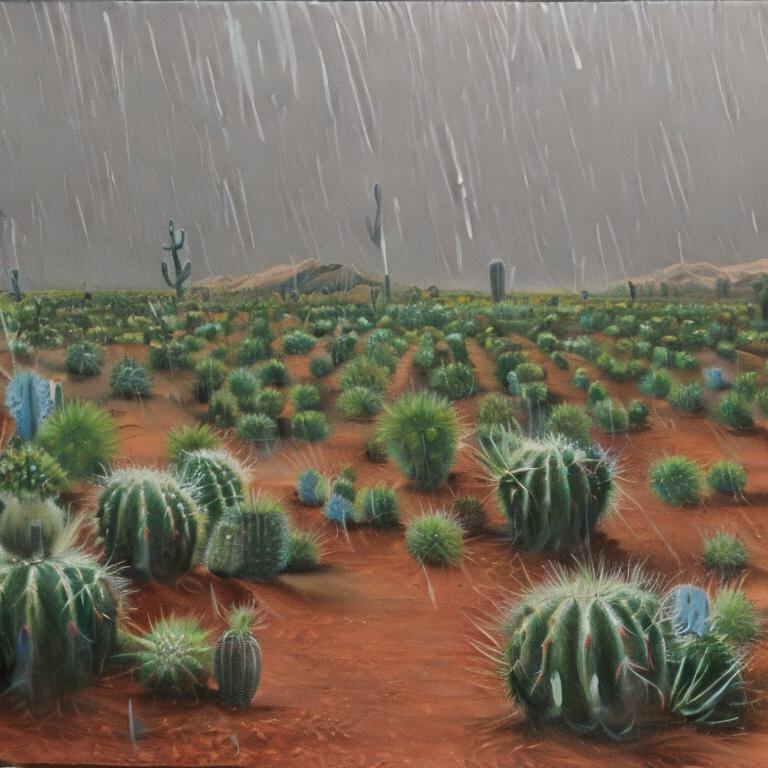}
         \caption{\responseref{}No adjustment.}
         \label{fig:attention_c_2}
     \end{subfigure}
     \hfill
     \begin{subfigure}[t]{0.3\linewidth}
         \centering
         \includegraphics[width=\textwidth]{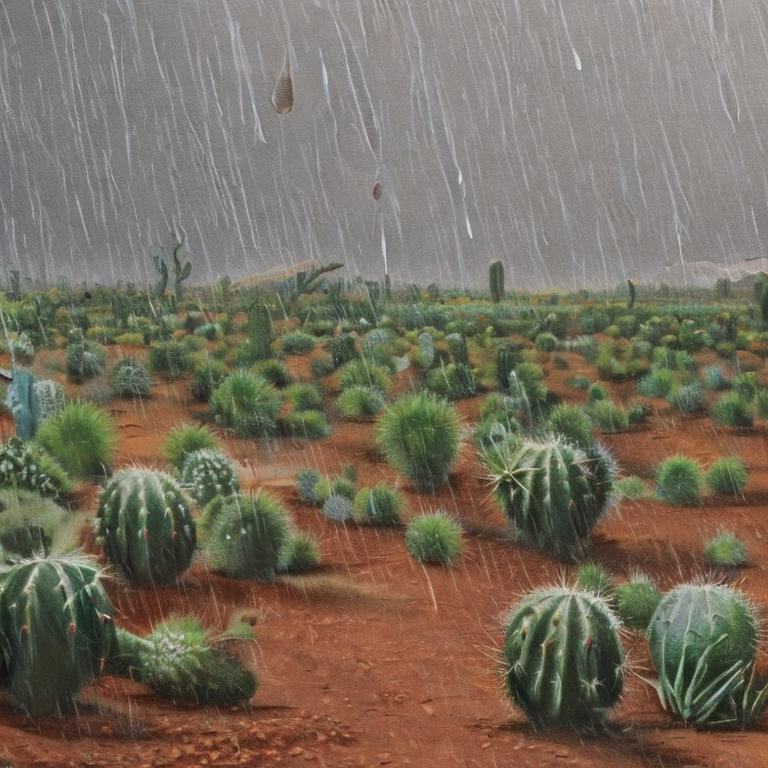}
         \caption{\responseref{}2x attention score to {\ttfamily [rain]}}
         \label{fig:attention_c_3}
     \end{subfigure}
    \caption{\responseref{}Adjusting model attention to the word {\ttfamily [rain]}.}
    \Description{This figure includes three different images. Image (a) has adjusted the model’s attention to [rain] to 0.5, and the generated image does not include rain drops. Image (b) does not include any model attention adjustment, and the generated image includes mild rain. Image (c) has adjusted the model’s attention to [rain] to 2, and the generated image includes heavy rain.}
    \label{fig:attention_exp_3}
\end{figure}

{\noindent \bf Example 4 prompt:}

{\ttfamily a painting of a lion, a [giraffe], a parrot, and a dolphin.} 

\begin{figure}[H]
     \centering
     \begin{subfigure}[t]{0.3\linewidth}
         \centering
         \includegraphics[width=\textwidth]{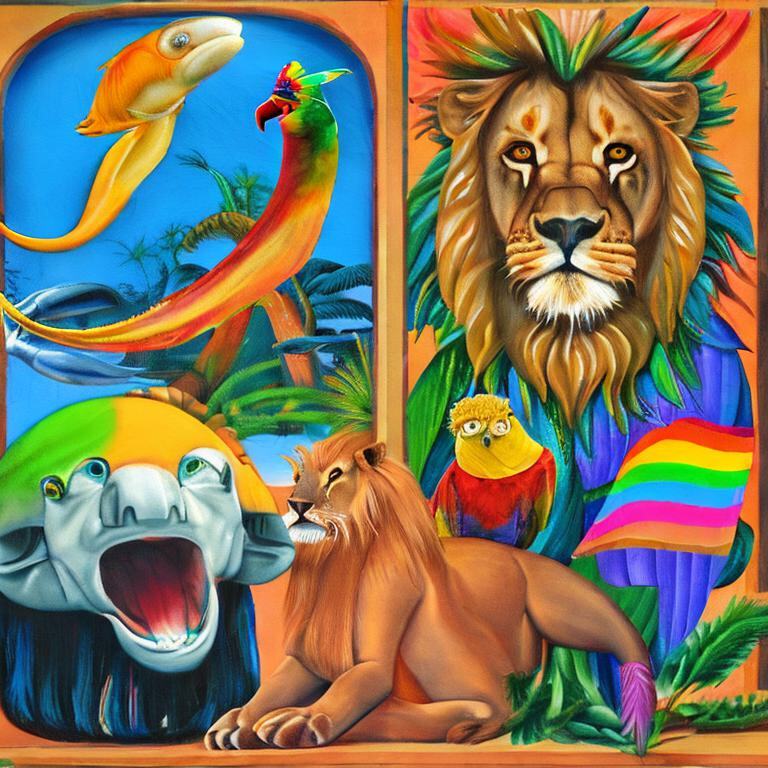}
         \caption{\responseref{}0.5x attention score to {\ttfamily [giraffe]}}
         \label{fig:attention_d_1}
     \end{subfigure}
     \hfill
     \begin{subfigure}[t]{0.3\linewidth}
         \centering
         \includegraphics[width=\textwidth]{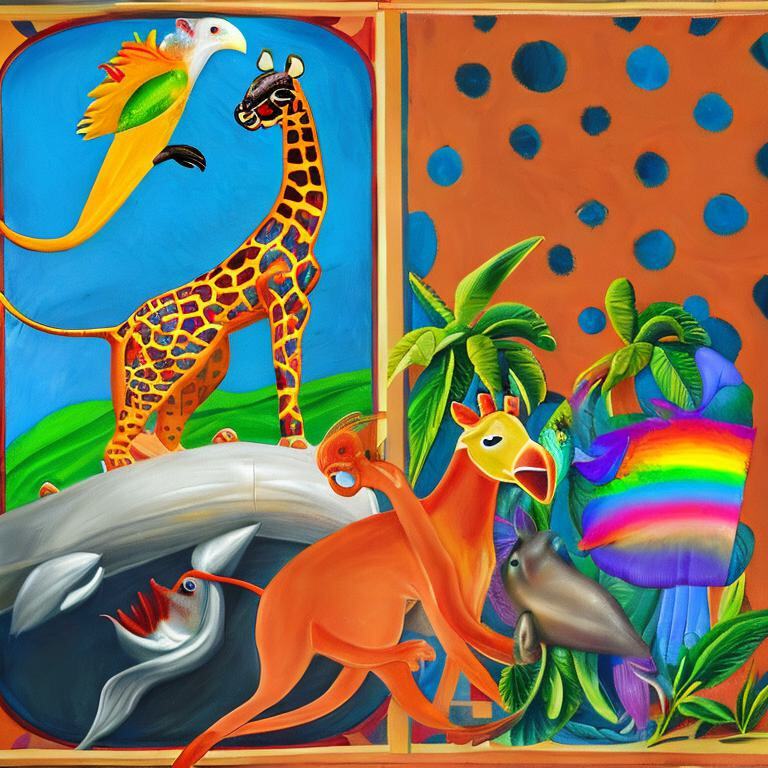}
         \caption{\responseref{}No adjustment.}
         \label{fig:attention_d_2}
     \end{subfigure}
     \hfill
     \begin{subfigure}[t]{0.3\linewidth}
         \centering
         \includegraphics[width=\textwidth]{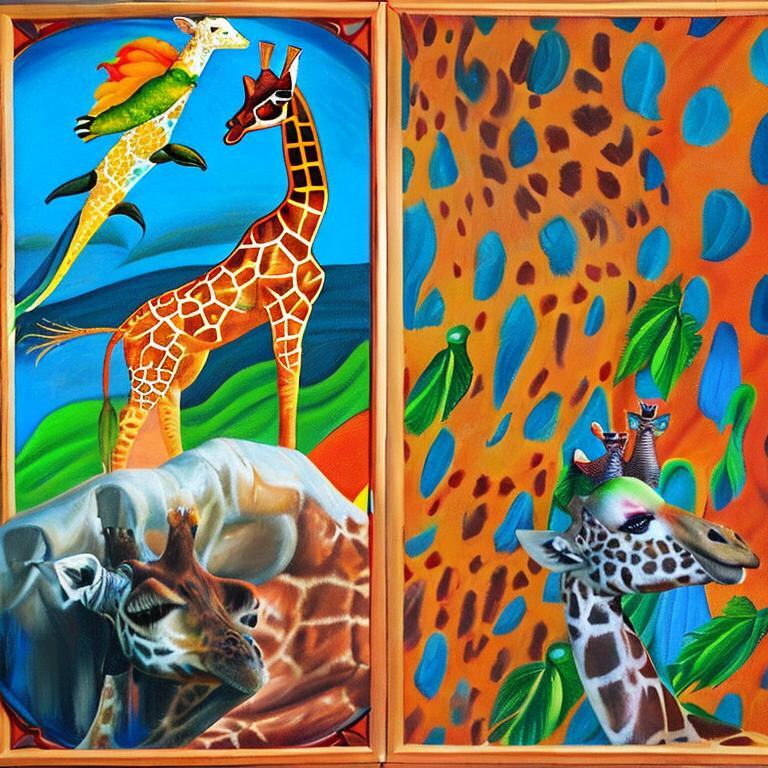}
         \caption{\responseref{}2x attention score to {\ttfamily [giraffe]}}
         \label{fig:attention_d_3}
     \end{subfigure}
    \caption{\responseref{}Adjusting model attention to the word {\ttfamily [giraffe]}.}
    \Description{This figure includes three different images. Image (a) has adjusted the model’s attention to [giraffe] to 0.5, and the generated image does not include any giraffes. Image (b) does not include any model attention adjustment, and the generated image includes one giraffe. Image (c) has adjusted the model’s attention to [giraffe] to 2, and the generated image includes multiple giraffes.}
    \label{fig:attention_exp_4}
\end{figure}

}

% \appendix









\end{document}